\documentclass[aps,prl,notitlepage,twocolumn]{revtex4-1} %linenumbers,endfloats
\newcommand{\w}{\columnwidth}
\usepackage[utf8]{inputenc}
\usepackage{siunitx}
\usepackage{graphicx}
\usepackage{amssymb}
\usepackage{hyperref}

\newcommand\pa{\textit{G\textsubscript{P1}}}
\newcommand\pd{\textit{G\textsubscript{P2}}}
\newcommand\ps{\textit{G\textsubscript{P3}}}
\newcommand\nl{\textit{G\textsubscript{N1}}}
\newcommand\nr{\textit{G\textsubscript{N2}}}
\newcommand\gd{\textit{G\textsubscript{D}}}
\newcommand{\lw}{\lambda_\mathrm{WG}}
\newcommand{\lno}{\lambda_\mathrm{Node}}
\newcommand{\f}[1]{\eta_\mathrm{Circuit{#1}}}
\newcommand{\fs}[1]{\eta_\mathrm{Source{#1}}}
\newcommand{\fw}[1]{\eta_\mathrm{WG{#1}}}
\newcommand{\fc}[1]{\eta_\mathrm{Capture{#1}}}
\newcommand{\fd}[1]{\eta_\mathrm{Detect{#1}}}
\newcommand{\fig}[1]{Fig. \ref{#1}}

\begin{document}

\title{Trapping and counting ballistic non-equilibrium electrons}

\author{Lars Freise}
\email{lars.freise@ptb.de}
\author{Thomas Gerster}
\author{David Reifert}
\author{Thomas Weimann}
\author{Klaus Pierz}
\author{Frank Hohls}
\author{Niels Ubbelohde}
\email[To whom correspondence should be addressed: ]{niels.ubbelohde@ptb.de}
\affiliation{Physikalisch-Technische Bundesanstalt, 38116 Braunschweig, Germany}

\date{February 4, 2020} %\today

\begin{abstract}
We demonstrate the trapping of electrons propagating ballistically at far-above-equilibrium energies in GaAs/AlGaAs heterostructures in high magnetic field.
We find low-loss transport along a gate-modified mesa edge in contrast to an effective decay of excess energy for the loop around a neighboring, mesa-confined node, enabling high-fidelity trapping.
Measuring the full counting statistics via single-charge detection yields the trapping (and escape) probabilities of electrons scattered (and excited) within the node.
Energetic and arrival-time distributions of captured electron wave packets are characterized by modulating tunnel barrier transmission.
\end{abstract}

\maketitle

The ability to prepare and subsequently detect discrete particles constitutes a crucial component for applications ranging from metrology over sensing to quantum information technologies, for example in utilizing single ions \cite{cirac1995QuantCompColdTrappedIon} or electrons \cite{imamoglu1999quantInfProcQdSpins} for quantum computation, or photons for quantum cryptography \cite{gisin2002quantCrypt}.
In electron quantum optics (EQO), the solid-state analogue to quantum optics, the recent introduction of on-demand single-electron sources is advancing experiments that have previously been realized with continuous electron sources \cite{henny1999fermionicHBT,oliver1999hbt,liu1998quantumInterference,ji2003electronicMachZehnder,neder2007twoParticleHBT,baeuerle2018reviewSEcontrol} by offering an inherent time-control, for instance in mesoscopic capacitors \cite{feve2007mesoscopicCapacitor,bocquillon2012hbt,bocquillon2013coherenceInditinguishabilityIndependentSources}, leviton injections \cite{dubois2013levitonEQO}, single-electron pumps (SEPs) \cite{blumenthal2007GHzQuantChargePump,fletcher2013clockControlledEmission_SEwavePackets,ubbelohde2015partitioningOnDemandElecPairs}, or the application of surface acoustic waves (SAWs) \cite{hermelin2011SawEqo,mcneil2011SingleElTransferDistantQDs}.
However, implementing a single-electron detector for an EQO experiment with ballistic electrons poses the challenge to detect ballistic electrons either on-the-fly or to trap them prior to detection.
Combining single-particle detection with an on-demand single-particle source provides direct access to the full counting statistics, independent of the bosonic or fermionic nature of the observed particles, and ensures high signal fidelities even in circuits with simultaneous multi-particle injection.
Additionally, multiple detectors can easily be combined and inherently record coincidence correlations.

\begin{figure}%[!tb]
	\includegraphics[width=\w]{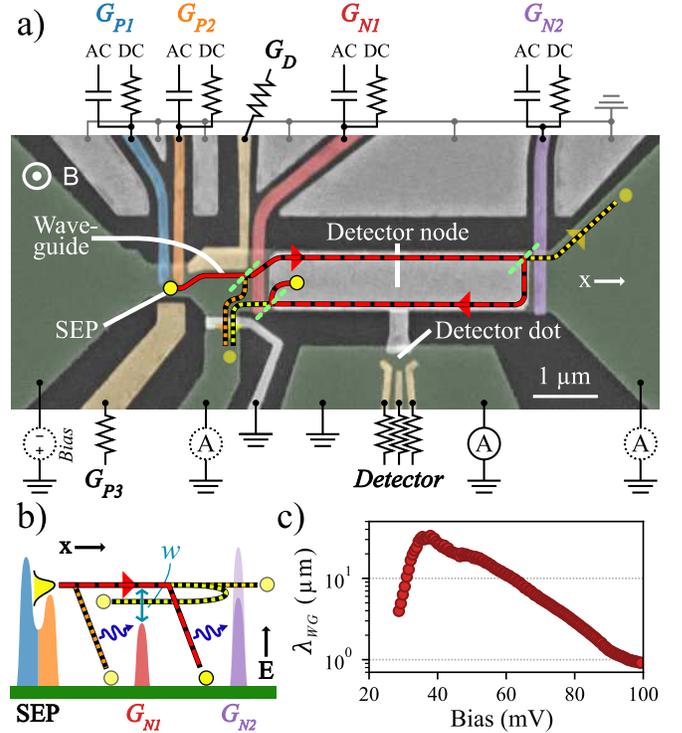}
	\caption{(a) False color SEM micrograph of sample A with measurement setup (dotted bias-source and ammeters exclusively employed for pre-characterization \cite{supp}). Active regions are colored (mesa green, DC-only gates yellow, DC-and-AC gates in different colors). Red/orange/yellow lines sketch possible transport paths for an electron (yellow dot), potential barrier transmissions of an electron are indicated as green beam splitters. (b) Sketched energy diagram showing the different propagation paths. (c) Bias-dependent scattering length $\lw$ (sample B, $V_\gd=\SI{0}{\milli\volt}$).}
	\label{pic:schematic}
\end{figure}

Here, we present a single-electron circuit that traps ballistic electrons emitted from an SEP-source at non-equilibrium energies of some ten \si{\milli\electronvolt}.
Continuous-current measurements characterize low-loss electron propagation towards a node, where single-electron counting provides a measure of the high scattering probability enabling trapping.
This technique resembles "sample and hold"-circuits in analog electronics, where a voltage is sampled by charging a capacitor that retains the voltage for readout.
The single-electron counterpart consists of four functional elements:
An electron is emitted on-demand \textit{(Source)} and ballistically traverses a waveguide section \textit{(WG)}. At the circuit’s output, the electron is then trapped inside a node by energy relaxation \textit{(Capture)} and read out \textit{(Detect)}.
This requires different propagation characteristics between waveguide (ideally lossless transport) and capture-node (ideally complete loss of excess energy) (\fig{pic:schematic}b), which will be analyzed throughout this article.
Following the denotation of detection efficiencies $\eta$ to register a photon at the detector, which is one of the key parameters in quantum optics setups \cite{hadfield2009singlePhotonDetectors}, separate coefficients $\fs{}$, $\fw{}$, $\fc{}$, and $\fd{}$ are assigned to the functional components, where $\left(1-\fw{}\right)$ and $\fc{}$ express the scattering probabilities in the waveguide and on the node, respectively, and $\f{}=\fs{}\fw{}\fc{}\fd{}$.

The two circuit geometries investigated (samples A and B, Figs. \ref{pic:schematic}a and S1 \cite{supp}, respectively) have been realized \cite{gerster2018qd} in the same GaAs/AlGaAs heterostructure with \SI{97}{\nano\meter} nominal 2DEG depth, charge carrier density $\SI{1.9e11}{\per\square\centi\meter}$, mobility $\SI{1.15e6}{\square\centi\meter\per\volt\per\second}$, and quantum life time $\SI{2.9}{\pico\second}$ \cite{coleridge1994resistivityPeakValues}. Components are formed by Cr/Au gates upon a shallow-etched channel. Measurements are carried out in a dry dilution refrigerator with base temperature below \SI{50}{\milli\kelvin} and at \SI{10}{\tesla} perpendicular magnetic field ($\nu\approx 1$).

A non-adiabatic single-parameter single-electron pump \cite{blumenthal2007GHzQuantChargePump,kaestner2008singleParamNonadQuantChargePump,maire2008noiseMeasSEP} is used as an on-demand source of non-equilibrium electrons.
It is confined by three gates (pump entrance and exit barriers \pa{} and \pd{}, side gate \ps{}) and excited by a $f=\SI{300}{\mega\hertz}$ sinusoidal pulse on \pa{}. The several-ten-\si{\milli\electronvolt} excess energy of the emitted electrons scales with \pd{}’s barrier height \cite{leicht2011SEPgenInQHedgestates,fletcher2013clockControlledEmission_SEwavePackets} and separates transport from the Fermi sea.
In continuous operation, this would correspond to a generated current $I=\si{\elementarycharge}f\approx\SI{48}{\pico\ampere}$ with elementary charge $\si{\elementarycharge}$.
SEPs have been reported in metrological applications with sub-ppm accuracy \cite{stein2015subppm} and hence do not currently limit the circuit fidelity. Here, $\fs{}\approx0.999$ is estimated from the difference of measured continuous current to $\si{\elementarycharge}f$ (not shown).

In perpendicular magnetic field, the injected electrons propagate along the sample edge with an imposed chirality, as in a ballistic waveguide. The confinement potential of this edge determines scattering cross sections and thereby the contributions of scattering mechanisms affecting propagating electrons.
The electrons’ injection energies substantially exceed those of systems in the IQHE $\nu=2$ regime at excess energies $\ll\SI{1}{\milli\electronvolt}$, where relaxation mechanics have extensively been elaborated (e.g. \cite{bocquillon2013coherenceInditinguishabilityIndependentSources,tewari2016quantumCoherenceAboreEF,nigg2016decoherenceHighEnElQHE,bocquillon2013review_EQOinBallisticConductors,kraehenmann2019augerSpectrSE_QHE,rodriguez2019stronEnergyRelaxQHE} and references therein).
In the waveguide section, the etch-defined edge potential is modulated by an edge-depletion gate \gd{} (here biased by $V_\gd=\SI{-350}{\milli\volt}$; stated voltages generally refer to signal’s DC parts), introduced for suppression of electron-electron interactions \cite{kataoka2016TOF_SEwavepacket_edgeSate} and shown to modulate and suppress LO phonon emission rates \cite{kataoka2016TOF_SEwavepacket_edgeSate,johnson2017phononEmissionRate} by minimizing scattering cross sections and spatially separating transport from the mesa edge.
Sample A is designed with a short waveguide length $l_\mathrm{WG}\approx\SI{1.25}{\micro\meter}$ to characterize scattering on the node unimpeded by residual losses in the waveguide.
The survival probability $P_\mathrm{S}$ of an electron ballistically reaching the detector node, equivalent to $\fw{}$, is obtained from the continuous current transmitted over the node entrance barrier \nl{} (not shown) as $P_\mathrm{S}=\fw{}\approx\frac{0.997}{\fs{}}\approx 0.998$.
A characteristic scattering length $\lambda$ can be estimated in a simple, exponential model \cite{palevski1989latTunneling+BallisTransp+Spectrosc2DEG,johnson2017phononEmissionRate,ota2019spectrosHotEl} from $P_\mathrm{S}=\exp{\left(-l/\lambda\right)}$, yielding $\lw\approx\SI{0.5}{\milli\meter}\gg l_\mathrm{WG}$, which expresses low-loss ballistic transport.

Sample B (Fig. S1 \cite{supp}) integrates more components and an increased waveguide length of $l_\mathrm{WG}\approx\SI{3}{\micro\meter}$ to demonstrate the functionality in more complex circuitry (data is marked explicitly).

Hot-electron measurements (as in \cite{ota2019spectrosHotEl,supp}) characterize $P_\mathrm{S}$ of ballistic electrons as a function of injection bias, yielding an energy-dependence of $\lw$ that is exemplarily depicted in \fig{pic:schematic}c \cite{fn6}:
At small injection energies $E$, electron-electron interactions dominate $\lw$ \mbox{\cite{ota2019spectrosHotEl,taubert2011phononRelaxation}} by
fully suppressing $P_\mathrm{S}$ (consistent with Ref. \cite{ota2019spectrosHotEl}, which describes the effect of electron-hole excitation by the injected hot electrons via a phenomenological, length-dependent threshold energy $E_\mathrm{th}$).
Maximal $\lw$ is attained above $E_\mathrm{th}$ (at few ten \si{\milli\volt}’s bias),
where the wave function overlap for electron-electron scattering is reduced and LO phonon emission not yet possible, before $\lw$ deteriorates with further increasing $E$, in accordance with enhancing LO phonon emission \cite{emary2016PhononEmission+ATD_SEP,emary2018phonon,supp}. 
Consequently, for EQO the injection energy of on-demand electrons should target the intermediate regime of optimal $\lw$.
In the waveguide section, $\lw$ is enhanced by modulating the edge potential via increasingly negative $V_\gd$, similar to observations in Ref. \cite{johnson2017phononEmissionRate}, however, it levels near $V_\gd\approx\SI{-300}{\milli\volt}$.
Extracting the confinement potential to enable calculation of predicted LO phonon emission rates \cite{kataoka2016TOF_SEwavepacket_edgeSate,johnson2017phononEmissionRate,emary2016PhononEmission+ATD_SEP} is inhibited by geometric restrictions preventing the required time-of-flight measurements. However, comparing $\lw$ to results in Ref. \cite{johnson2017phononEmissionRate} indicates phonon emission rates similarly exceeding model predictions \cite{fn5}, substantiating the presence of additional processes, possibly indirect LO phonon emission involving supplemental LA phonon emission \cite{johnson2017phononEmissionRate,emary2018phonon}.

Electrons that scattered while propagating through the waveguide are reflected by \nl{} and sunk into an ohmic side contact.
Propagation across the node in an etch-defined mesa-edge potential is investigated by recording $\fc{}$ of trapped electrons within a counting scheme, yielding a combined scattering probability of all mechanisms contributing to trapping.
A distance of $\SI{4.2}{\micro\meter}$ between \nl{} and node exit barrier \nr{} creates a large node extent, forming
an elongated quantum dot with lithographic circumference of $\SI{10.4}{\micro\meter}$ and typical charging energy of few ten \si{\micro\electronvolt}. Assuming drift velocities of magnitude $\SI{5e4}{\meter\per\second}$  \cite{kataoka2016TOF_SEwavepacket_edgeSate,johnson2017phononEmissionRate} would imply node transit times below \SI{100}{\pico\second}, distinctly exceeding reported temporal wave packet extents of few ten picoseconds \cite{waldie2015elWavePacketsSEP,kataoka2017timeResolvedWPdetection,johnson2016ultrafastWavePackSampling}.

\begin{figure}%[!tb]
	\includegraphics[width=\w]{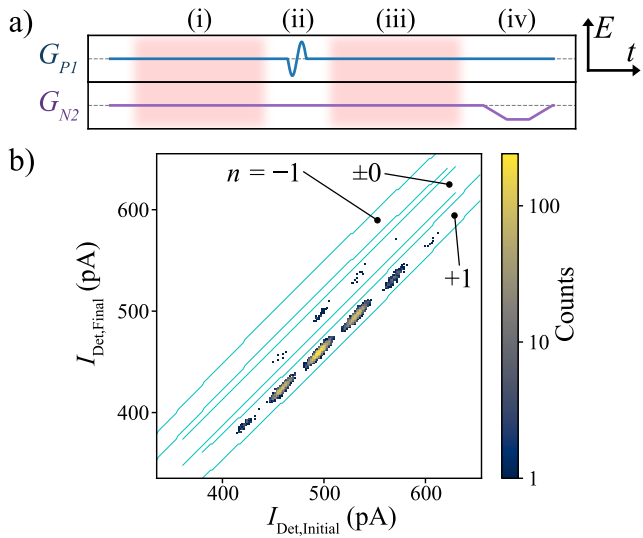}
	\caption{(a) Sketched sequence of AC modulations built from four intervals [initial (i)/final (iii) detector state readout (marked in red), (ii) electron transport, (iv) reset; not to scale] used for each measurement cycle. (b) Exemplary counting histogram ($V_\nl=\SI{-334}{\milli\volt}$ in \fig{pic:EnSpec}a). Measured detector signal is categorized into 200 bins and shown as final over initial detector state. Cyan lines enclose counts identifying a change by $-1$, $0$, or $+1$ electrons.}
	\label{pic:hist}
\end{figure}

The detector dot is formed by split-gates against the edge of a separate mesa structure, an additional floating gate increases capacitive coupling to the node.
Separation of the two mesas avoids the strong electrostatic screening and resulting reduction in detector resolution accompanying complete split-gate realizations \cite{roessler2010hybridQDchargeDet}.
The detector dot is operated in the Coulomb blockade regime to record charge changes on the node, with two readouts per repetition of the counting cycle, following the sequence depicted in \fig{pic:hist}a:
after completing the first readout [of the initial detector state, (i)], emission of an electron is triggered (ii), while after the second readout [final detector state, (iii)], the node state is reset (iv).
The cycle repetition rate of \SI{24}{\hertz} is dominated by the charge-detection bandwidth of detector dot readout (\SI{62.5}{\hertz} to \SI{83.3}{\hertz}) and is limited by the coupling of an added electron onto the detector dot (which was observed to decrease with increasing node size and incomplete floating-gate coverage), the detector dot sensitivity and the required signal-to-noise ratio.

To estimate $\fc{}$ and $\fd{}$, $10^5$ cycles are recorded and the data binned into a 2D histogram, depicting final over initial detector state. Exemplary data can be seen in \fig{pic:hist}b, where counts on the diagonal represent constant detector signals (i.e. node’s charge state remains unchanged within a given cycle), while off-diagonal signatures reflect changes in the deposited charge.
Multiple clusters emerge, clearly separated in x-direction [corresponding to a varying node occupation in interval (i) due to the reset (iv)] and in y-direction [corresponding to electrons emitted on-demand in interval (ii)].
Despite the large node size of $>4\,\si{\micro\meter\squared}$, a peak separation of $\geq5\sigma$ \cite{supp} is achieved. With $1-\fd{}\lessapprox 10^{-6}$, identification errors do not limit $\f{}$.
The counts are summed to $N_n$ (cyan lines in \fig{pic:hist}b) and lead to the full counting statistics $P_n=\frac{N_n}{\sum_iN_i}$ of a change in the node’s charge by $n$ electrons.
The high probability $P_1\approx0.996$ indicates efficient scattering of electrons in this segment and is equivalent to the achieved maximal circuit fidelity, $\f{}\approx 0.996$, revealing the near-unity $\fc{}=\frac{\f{}}{\fs{}\fw{}\fd{}}\approx0.999$ that is similarly reproduced in sample B (cf. \cite{supp}).
The achievable fidelity of these circuits is largely limited by transport properties of the waveguide and compares well to fidelities of (non-ballistic) electron transport inside SAW-induced moving quantum dots \cite{hermelin2011SawEqo,takada2019sawBeamsplitter}.

\begin{figure}%[htb]
\includegraphics[width=\w]{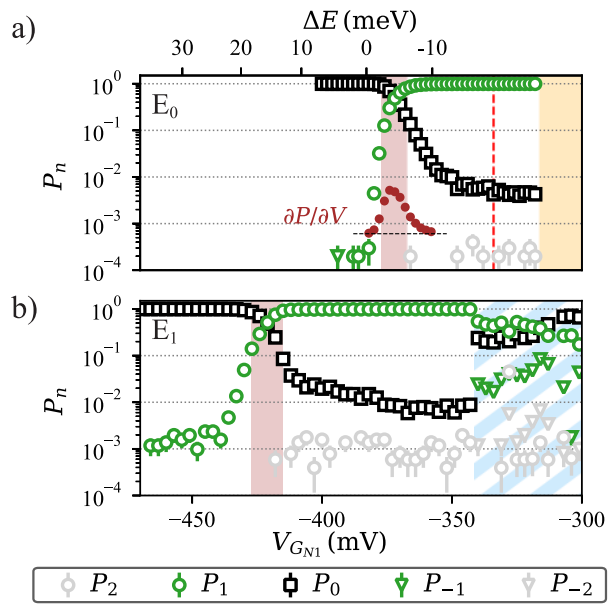}
\caption{Energy-dependent probabilities for detection of single electron wave packets depending on node entrance barrier heights for two electron emission energies $E_0$ (a) and $E_1=E_0+\SI{20}{\milli\electronvolt}$ (b) plotted on identical scales (\nr{} raised above emission energies). Barrier height increases to the left, transitions $P_1\approx1 \rightarrow P_1\approx0$ depict energy distributions of wave packets \cite{fn34}. FWHMs are highlighted in dark red, (a) with inset of derivative $\frac{\partial P_1}{\partial V_\nl{}}$. Error bars represent statistical errors. Orange background and blue stripes mark regions inaccessible to detector readout (no clear, consistent charge state observable, hinting at non-linear transmission properties of \nl{}) and of missing electrons, respectively. Red line indicates the value taken for \fig{pic:special}a. $P_2>0$ due to non-ideal SEP-quantization.}
\label{pic:EnSpec}
\end{figure}

For the trapping of on-demand electron wave packets, it is particularly important to also consider the intrinsic energy-time correlation of the wave packets emitted by SEPs \cite{slava2017classQuantCrossover,fletcher2019tomography}.
Within our trap-and-detect scheme, these wave packet characteristics are in the following accessed by utilizing the energy-dependent transmission of the node-defining barriers \cite{fletcher2013clockControlledEmission_SEwavePackets,waldie2015elWavePacketsSEP,kataoka2016TOF_SEwavepacket_edgeSate,ubbelohde2015partitioningOnDemandElecPairs,johnson2016ultrafastWavePackSampling,fletcher2019tomography}. The resulting tomographic projections of the wave packet furthermore provide the basis for the synchronization of energetic, temporal, and spatial wave packet overlaps required for interferences of electron wave packets in non-equilibrium EQO experiments.

Figure \ref{pic:EnSpec}a demonstrates the progressive blocking of lower-energy electrons with increasing barrier height of \nl{}, depicting the wave packet’s energy distribution, with an asymmetry (cf. inset) caused by non-Gaussian emission characteristics of the SEP \cite{slava2017classQuantCrossover,fletcher2019tomography}.
A relative energy-scale can be derived \cite{supp} and is displayed on the top axis of \fig{pic:EnSpec}a (zero set for maximum electron energy arriving at \nl{}).
The energetic width of the distribution is evaluated as the transition’s full width half maximum (FWHM) in $\frac{\partial P_n}{\partial V_\nl{}}$ to $\delta E\approx \SI{4}{\milli\electronvolt}$ and agrees well with other experiments \cite{fletcher2013clockControlledEmission_SEwavePackets,waldie2015elWavePacketsSEP,johnson2016ultrafastWavePackSampling}. This provides an upper bound to the energetic width of the wave packet \cite{waldie2015elWavePacketsSEP} that is fundamentally limited by convolution with the barrier characteristics, the relative contribution of which is difficult to estimate even in a recent tomography scheme \cite{fletcher2019tomography}.

The SEP’s emission energy $E_0$ considered so far within \fig{pic:EnSpec}a lies in the maximal-$\lw{}$ region, close to the gradual onset of LO phonon emission \cite{supp}.
However, the emission energies of SEPs can span a large range, exceeding $E>\SI{100}{\milli\electronvolt}$ \cite{fletcher2013clockControlledEmission_SEwavePackets,leicht2011SEPgenInQHedgestates}.
Trapping at an increased emission energy $E_1$ is shown in \fig{pic:EnSpec}b ($\Delta E_1=E_1-E_0\approx \SI{20}{\milli\electronvolt}$ \cite{supp}, compare positions of both transitions in \fig{pic:EnSpec}; energy increase by synchronized excitation of \pa{} and \pd{}) and exhibits consistently high $\fc{}^{E_1}\approx\fc{}^{E_0}$ despite the increased emission energy, as apparent from $\f{}^{E_1}\approx 0.992$.
The energetic width $\delta E^{E_1}\approx \SI{5}{\milli\electronvolt}$ is also close to the lower-energy value.
A larger energy increase by $\Delta E_2\approx 50\,\mathrm{meV}$ ($E_2=E_0+\Delta E_2$, data in \cite{supp}) yields $\delta E^{E_2}\approx\SI{9.5}{\milli\electronvolt}$ and reduced $\f{}^{E_2}\approx 0.90$, which however is mainly due to deteriorating quantization of the source, $\fs{}^{E_2}$. There is no discernible indication for a decrease in $\fc{}$ over the inspected \SI{50}{\milli\electronvolt} emission-energy range.

In the energy-resolved measurements at increased emission energies, additional signatures appear in the limit of small node entrance barrier heights (blue stripes in \fig{pic:EnSpec}b, \cite{fn2}), where $P_1$ drops for decreasing barrier heights, while $P_0$ increases and in some cycles the charge on the node is reduced, $P_{-1}>P_{-2}>0$.
The appearing $P_{-1},P_{-2}>0$ cannot be plausibly explained as a modulation of source-, waveguide-, or detector-functionality and is directly related to electrons emitted from the SEP \cite{fn1}. We interpret $P_{-1},P_{-2}>0$ as a process of energy-redistribution from relaxation of the hot electron, similar to observations in Ref. \cite{taubert2010avalancheAmplifier}, enabling thereby excited electrons to escape the node.
At $E_2$, this region extends towards stronger negative $V_\nl$, in accordance with the increased energy available for redistribution upon scattering.
Including $P_{-2}$, four non-zero elements of the full counting statistics are recorded within the low-barrier limit of \fig{pic:EnSpec}b, highlighting the benefit of the single-electron resolution.

\begin{figure}%[htb]
	\includegraphics[width=\w]{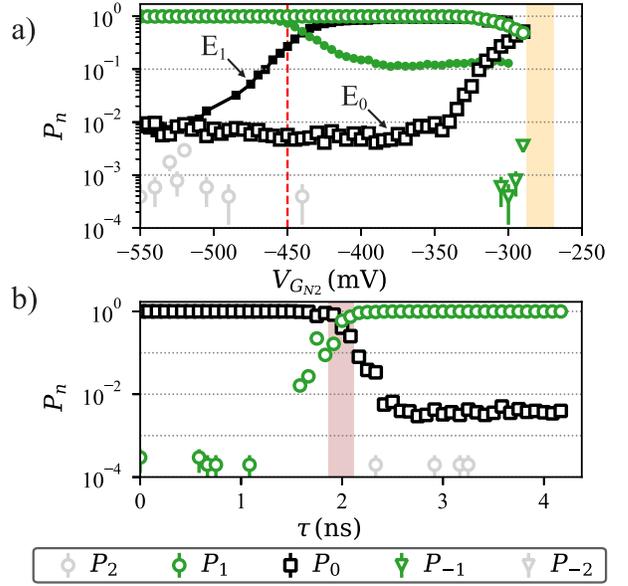}
	\caption{(a) Energy-resolved capturing for \nr{} probes scattering while traversing the node. Data shown at $E_0$ ($V_\nl^{E_0}=\SI{-334}{\milli\volt}$) and $E_1$ (small markers; for clarity only $P_0$ and $P_1$ shown, $V_\nl^{E_1}=\SI{-370}{\milli\volt}$). Opposed to \fig{pic:EnSpec}, transmission over \nr{} results in $n=0$. Red line marks value taken for \fig{pic:EnSpec}a. (b) Time-resolved capturing at $E_0$ by raising \nl{} at time delay $\tau$ relative to pump excitation (methods in \cite{supp}).}
	\label{pic:special}
\end{figure}

The wave packet’s energy distributions at \nr{} are depicted in \fig{pic:special}a, probing scattering while traversing the node.
%Scattering while traversing the node is probed by scanning \nr{} (\fig{pic:special}a).
At $E_0$, no survival probability of electrons ballistically traversing the node is resolved [$P_\mathrm{S}^{E_0}\approx 0$, visible from (incomplete) wave packet mapping only at vanishing barrier heights].
Thus, merely by taking into account scattered electrons maintaining sufficient energy to pass small \nr{}, $P_\mathrm{Max.\,transmission}\approx 0.52$ can be used to estimate an upper limit for $\lno^{E_0}\ll\SI{6.5}{\micro\meter}$. This represents, in contrast to propagation in the gate-modified potential of the waveguide section, a change of approximately two orders of magnitude for transport through the etch-defined edge potential and for $E_0$ close to the energetic regime of strong electron-electron scattering (cf. \cite{supp}).

At the increased emission energy $E_1$, however, a substantial ballistic survival probability $P_\mathrm{S}^{E_1}\approx 0.88$ is evident from the clear transition detected at elevated barrier \nr{} (also \fig{pic:special}a), corresponding to $\lno^{E_1}\approx\SI{34}{\micro\meter}$.
$\lno^{E_1} > l_\mathrm{Node\,circumference}$ would suggest a pronounced fraction of electrons being able to escape the node after traversing its circumference in a full loop. This is however contradicted by the miniscule escape probabilities seen in the low-barrier limits of \fig{pic:EnSpec}, which indicate an even further reduced scattering length with near-unity $\fc{}$’s.
We conclude that more mechanisms beyond the changing edge potential are contributing to these high trapping fidelities, for example backactions of the active charge detector \cite{harbusch2010detectorBackaction}. The combined scattering probability, however, methodically cannot identify individual scattering mechanisms.
Furthermore, additional modifications of scattering may result from the node being isolated, e.g. by loop excitations ($E_\mathrm{Loop\,excitation}\ll E_0$), as for example observed in Ref. \cite{altimiras2010tuningEnergyRelax}.
Manipulating the node’s electrochemical potential or, via in-plane gating, its edge-potential profile did not measurably enhance $\fc{}$.
Ultimately, defining a narrow, energy-selective "capture window" (labeled $w$ in \fig{pic:schematic}b) would allow detecting energetically sharp, incident electron wave packets even after minor energy losses.

The wave packet’s arrival-time distribution at \nl{} is pictured in \fig{pic:special}b as a function of the time delay $\tau$ at which \nl{} is raised to block the incident wave packets from entering the node (cf. \cite{supp}).
Its relatively wide FWHM $\delta t\approx\SI{250}{\pico\second}$ is an upper bound to the temporal width of the electron wave packet \cite{supp}, limited by barrier characteristics but consistent with the dependence on the SEP’s operating point \cite{fletcher2013clockControlledEmission_SEwavePackets,fletcher2019tomography,slava2017classQuantCrossover}. Despite $\delta t$ being of the same order of magnitude as the estimated transit time around the node, the distinct trapping capability persists due to the high scattering probability on the node preventing electrons from exiting after fully traversing its circumference.
This technique of time-resolved detection also constitutes the minimal module for a proposed single-shot voltage sampling \cite{johnson2016ultrafastWavePackSampling}.

In conclusion, strongly differing scattering probabilities were measured in neighboring segments of a single-electron circuit, allowing capturing and detection of single electrons after ballistic propagation, resolving the full counting statistics and estimating wave packet characteristics. The resolution of these can prospectively be enhanced by manipulating SEP emission characteristics and applying tomographic schemes \cite{fletcher2019tomography} or optimizing barrier characteristics \cite{locane2019dynScattSEWP}.
The demonstrated robustness of detection to shifting injection energies is especially important for future collision and interferometry experiments requiring an energetic overlap between wave packets emitted from independent non-equilibrium sources.

\begin{acknowledgments}
We gratefully acknowledge MBE operation by H. Marx and discussions with D. Maradan and V. Kashcheyevs.
This work was supported in part by the Joint Research Project ‘SEQUOIA’ (17FUN04). This project has received funding from the European Metrology Programme for Innovation and Research (EMPIR) co-financed by the Participating States and from the European Union’s Horizon 2020 research and innovation programme. We additionally acknowledge support by DFG within cluster of excellence 'QuantumFrontiers'.
\end{acknowledgments}

\bibliographystyle{apsrev4-2}
\bibliography{literature}

\begin{thebibliography}{54}%
\makeatletter
\providecommand \@ifxundefined [1]{%
 \@ifx{#1\undefined}
}%
\providecommand \@ifnum [1]{%
 \ifnum #1\expandafter \@firstoftwo
 \else \expandafter \@secondoftwo
 \fi
}%
\providecommand \@ifx [1]{%
 \ifx #1\expandafter \@firstoftwo
 \else \expandafter \@secondoftwo
 \fi
}%
\providecommand \natexlab [1]{#1}%
\providecommand \enquote  [1]{``#1''}%
\providecommand \bibnamefont  [1]{#1}%
\providecommand \bibfnamefont [1]{#1}%
\providecommand \citenamefont [1]{#1}%
\providecommand \href@noop [0]{\@secondoftwo}%
\providecommand \href [0]{\begingroup \@sanitize@url \@href}%
\providecommand \@href[1]{\@@startlink{#1}\@@href}%
\providecommand \@@href[1]{\endgroup#1\@@endlink}%
\providecommand \@sanitize@url [0]{\catcode `\\12\catcode `\$12\catcode
  `\&12\catcode `\#12\catcode `\^12\catcode `\_12\catcode `\%12\relax}%
\providecommand \@@startlink[1]{}%
\providecommand \@@endlink[0]{}%
\providecommand \url  [0]{\begingroup\@sanitize@url \@url }%
\providecommand \@url [1]{\endgroup\@href {#1}{\urlprefix }}%
\providecommand \urlprefix  [0]{URL }%
\providecommand \Eprint [0]{\href }%
\providecommand \doibase [0]{https://doi.org/}%
\providecommand \selectlanguage [0]{\@gobble}%
\providecommand \bibinfo  [0]{\@secondoftwo}%
\providecommand \bibfield  [0]{\@secondoftwo}%
\providecommand \translation [1]{[#1]}%
\providecommand \BibitemOpen [0]{}%
\providecommand \bibitemStop [0]{}%
\providecommand \bibitemNoStop [0]{.\EOS\space}%
\providecommand \EOS [0]{\spacefactor3000\relax}%
\providecommand \BibitemShut  [1]{\csname bibitem#1\endcsname}%
\let\auto@bib@innerbib\@empty
%</preamble>
\bibitem [{\citenamefont {Cirac}\ and\ \citenamefont
  {Zoller}(1995)}]{cirac1995QuantCompColdTrappedIon}%
  \BibitemOpen
  \bibfield  {author} {\bibinfo {author} {\bibfnamefont {J.~I.}\ \bibnamefont
  {Cirac}}\ and\ \bibinfo {author} {\bibfnamefont {P.}~\bibnamefont {Zoller}},\
  }\href {https://doi.org/10.1103/PhysRevLett.74.4091} {\bibfield  {journal}
  {\bibinfo  {journal} {Phys. Rev. Lett.}\ }\textbf {\bibinfo {volume} {74}},\
  \bibinfo {pages} {4091} (\bibinfo {year} {1995})}\BibitemShut {NoStop}%
\bibitem [{\citenamefont {Imamoglu}\ \emph {et~al.}(1999)\citenamefont
  {Imamoglu}, \citenamefont {Awschalom}, \citenamefont {Burkard}, \citenamefont
  {DiVincenzo}, \citenamefont {Loss}, \citenamefont {Sherwin},\ and\
  \citenamefont {Small}}]{imamoglu1999quantInfProcQdSpins}%
  \BibitemOpen
  \bibfield  {author} {\bibinfo {author} {\bibfnamefont {A.}~\bibnamefont
  {Imamoglu}}, \bibinfo {author} {\bibfnamefont {D.~D.}\ \bibnamefont
  {Awschalom}}, \bibinfo {author} {\bibfnamefont {G.}~\bibnamefont {Burkard}},
  \bibinfo {author} {\bibfnamefont {D.~P.}\ \bibnamefont {DiVincenzo}},
  \bibinfo {author} {\bibfnamefont {D.}~\bibnamefont {Loss}}, \bibinfo {author}
  {\bibfnamefont {M.}~\bibnamefont {Sherwin}},\ and\ \bibinfo {author}
  {\bibfnamefont {A.}~\bibnamefont {Small}},\ }\href
  {https://doi.org/10.1103/PhysRevLett.83.4204} {\bibfield  {journal} {\bibinfo
   {journal} {Phys. Rev. Lett.}\ }\textbf {\bibinfo {volume} {83}},\ \bibinfo
  {pages} {4204} (\bibinfo {year} {1999})}\BibitemShut {NoStop}%
\bibitem [{\citenamefont {Gisin}\ \emph {et~al.}(2002)\citenamefont {Gisin},
  \citenamefont {Ribordy}, \citenamefont {Tittel},\ and\ \citenamefont
  {Zbinden}}]{gisin2002quantCrypt}%
  \BibitemOpen
  \bibfield  {author} {\bibinfo {author} {\bibfnamefont {N.}~\bibnamefont
  {Gisin}}, \bibinfo {author} {\bibfnamefont {G.}~\bibnamefont {Ribordy}},
  \bibinfo {author} {\bibfnamefont {W.}~\bibnamefont {Tittel}},\ and\ \bibinfo
  {author} {\bibfnamefont {H.}~\bibnamefont {Zbinden}},\ }\href
  {https://doi.org/10.1103/RevModPhys.74.145} {\bibfield  {journal} {\bibinfo
  {journal} {Rev. Mod. Phys.}\ }\textbf {\bibinfo {volume} {74}},\ \bibinfo
  {pages} {145} (\bibinfo {year} {2002})}\BibitemShut {NoStop}%
\bibitem [{\citenamefont {Henny}\ \emph {et~al.}(1999)\citenamefont {Henny},
  \citenamefont {Oberholzer}, \citenamefont {Strunk}, \citenamefont {Heinzel},
  \citenamefont {Ensslin}, \citenamefont {Holland},\ and\ \citenamefont
  {Sch{\"o}nenberger}}]{henny1999fermionicHBT}%
  \BibitemOpen
  \bibfield  {author} {\bibinfo {author} {\bibfnamefont {M.}~\bibnamefont
  {Henny}}, \bibinfo {author} {\bibfnamefont {S.}~\bibnamefont {Oberholzer}},
  \bibinfo {author} {\bibfnamefont {C.}~\bibnamefont {Strunk}}, \bibinfo
  {author} {\bibfnamefont {T.}~\bibnamefont {Heinzel}}, \bibinfo {author}
  {\bibfnamefont {K.}~\bibnamefont {Ensslin}}, \bibinfo {author} {\bibfnamefont
  {M.}~\bibnamefont {Holland}},\ and\ \bibinfo {author} {\bibfnamefont
  {C.}~\bibnamefont {Sch{\"o}nenberger}},\ }\href
  {https://doi.org/10.1126/science.284.5412.296} {\bibfield  {journal}
  {\bibinfo  {journal} {Science}\ }\textbf {\bibinfo {volume} {284}},\ \bibinfo
  {pages} {296} (\bibinfo {year} {1999})}\BibitemShut {NoStop}%
\bibitem [{\citenamefont {Oliver}\ \emph {et~al.}(1999)\citenamefont {Oliver},
  \citenamefont {Kim}, \citenamefont {Liu},\ and\ \citenamefont
  {Yamamoto}}]{oliver1999hbt}%
  \BibitemOpen
  \bibfield  {author} {\bibinfo {author} {\bibfnamefont {W.~D.}\ \bibnamefont
  {Oliver}}, \bibinfo {author} {\bibfnamefont {J.}~\bibnamefont {Kim}},
  \bibinfo {author} {\bibfnamefont {R.~C.}\ \bibnamefont {Liu}},\ and\ \bibinfo
  {author} {\bibfnamefont {Y.}~\bibnamefont {Yamamoto}},\ }\href
  {https://doi.org/10.1126/science.284.5412.299} {\bibfield  {journal}
  {\bibinfo  {journal} {Science}\ }\textbf {\bibinfo {volume} {284}},\ \bibinfo
  {pages} {299} (\bibinfo {year} {1999})}\BibitemShut {NoStop}%
\bibitem [{\citenamefont {Liu}\ \emph {et~al.}(1998)\citenamefont {Liu},
  \citenamefont {Odom}, \citenamefont {Yamamoto},\ and\ \citenamefont
  {Tarucha}}]{liu1998quantumInterference}%
  \BibitemOpen
  \bibfield  {author} {\bibinfo {author} {\bibfnamefont {R.~C.}\ \bibnamefont
  {Liu}}, \bibinfo {author} {\bibfnamefont {B.}~\bibnamefont {Odom}}, \bibinfo
  {author} {\bibfnamefont {Y.}~\bibnamefont {Yamamoto}},\ and\ \bibinfo
  {author} {\bibfnamefont {S.}~\bibnamefont {Tarucha}},\ }\href
  {https://doi.org/10.1038/34611} {\bibfield  {journal} {\bibinfo  {journal}
  {Nature}\ }\textbf {\bibinfo {volume} {391}},\ \bibinfo {pages} {263}
  (\bibinfo {year} {1998})}\BibitemShut {NoStop}%
\bibitem [{\citenamefont {Ji}\ \emph {et~al.}(2003)\citenamefont {Ji},
  \citenamefont {Chung}, \citenamefont {Sprinzak}, \citenamefont {Heiblum},
  \citenamefont {Mahalu},\ and\ \citenamefont
  {Shtrikman}}]{ji2003electronicMachZehnder}%
  \BibitemOpen
  \bibfield  {author} {\bibinfo {author} {\bibfnamefont {Y.}~\bibnamefont
  {Ji}}, \bibinfo {author} {\bibfnamefont {Y.}~\bibnamefont {Chung}}, \bibinfo
  {author} {\bibfnamefont {D.}~\bibnamefont {Sprinzak}}, \bibinfo {author}
  {\bibfnamefont {M.}~\bibnamefont {Heiblum}}, \bibinfo {author} {\bibfnamefont
  {D.}~\bibnamefont {Mahalu}},\ and\ \bibinfo {author} {\bibfnamefont
  {H.}~\bibnamefont {Shtrikman}},\ }\href {https://doi.org/10.1038/nature01503}
  {\bibfield  {journal} {\bibinfo  {journal} {Nature}\ }\textbf {\bibinfo
  {volume} {422}},\ \bibinfo {pages} {415} (\bibinfo {year}
  {2003})}\BibitemShut {NoStop}%
\bibitem [{\citenamefont {Neder}\ \emph {et~al.}(2007)\citenamefont {Neder},
  \citenamefont {Ofek}, \citenamefont {Chung}, \citenamefont {Heiblum},
  \citenamefont {Mahalu},\ and\ \citenamefont
  {Umansky}}]{neder2007twoParticleHBT}%
  \BibitemOpen
  \bibfield  {author} {\bibinfo {author} {\bibfnamefont {I.}~\bibnamefont
  {Neder}}, \bibinfo {author} {\bibfnamefont {N.}~\bibnamefont {Ofek}},
  \bibinfo {author} {\bibfnamefont {Y.}~\bibnamefont {Chung}}, \bibinfo
  {author} {\bibfnamefont {M.}~\bibnamefont {Heiblum}}, \bibinfo {author}
  {\bibfnamefont {D.}~\bibnamefont {Mahalu}},\ and\ \bibinfo {author}
  {\bibfnamefont {V.}~\bibnamefont {Umansky}},\ }\href
  {https://doi.org/10.1038/nature05955} {\bibfield  {journal} {\bibinfo
  {journal} {Nature}\ }\textbf {\bibinfo {volume} {448}},\ \bibinfo {pages}
  {333} (\bibinfo {year} {2007})}\BibitemShut {NoStop}%
\bibitem [{\citenamefont {Bäuerle}\ \emph {et~al.}(2018)\citenamefont
  {Bäuerle}, \citenamefont {Glattli}, \citenamefont {Meunier}, \citenamefont
  {Portier}, \citenamefont {Roche}, \citenamefont {Roulleau}, \citenamefont
  {Takada},\ and\ \citenamefont {Waintal}}]{baeuerle2018reviewSEcontrol}%
  \BibitemOpen
  \bibfield  {author} {\bibinfo {author} {\bibfnamefont {C.}~\bibnamefont
  {Bäuerle}}, \bibinfo {author} {\bibfnamefont {D.~C.}\ \bibnamefont
  {Glattli}}, \bibinfo {author} {\bibfnamefont {T.}~\bibnamefont {Meunier}},
  \bibinfo {author} {\bibfnamefont {F.}~\bibnamefont {Portier}}, \bibinfo
  {author} {\bibfnamefont {P.}~\bibnamefont {Roche}}, \bibinfo {author}
  {\bibfnamefont {P.}~\bibnamefont {Roulleau}}, \bibinfo {author}
  {\bibfnamefont {S.}~\bibnamefont {Takada}},\ and\ \bibinfo {author}
  {\bibfnamefont {X.}~\bibnamefont {Waintal}},\ }\href
  {https://doi.org/10.1088/1361-6633/aaa98a} {\bibfield  {journal} {\bibinfo
  {journal} {Reports on Progress in Physics}\ }\textbf {\bibinfo {volume}
  {81}},\ \bibinfo {pages} {056503} (\bibinfo {year} {2018})}\BibitemShut
  {NoStop}%
\bibitem [{\citenamefont {F{\`e}ve}\ \emph {et~al.}(2007)\citenamefont
  {F{\`e}ve}, \citenamefont {Mah{\'e}}, \citenamefont {Berroir}, \citenamefont
  {Kontos}, \citenamefont {Pla{\c c}ais}, \citenamefont {Glattli},
  \citenamefont {Cavanna}, \citenamefont {Etienne},\ and\ \citenamefont
  {Jin}}]{feve2007mesoscopicCapacitor}%
  \BibitemOpen
  \bibfield  {author} {\bibinfo {author} {\bibfnamefont {G.}~\bibnamefont
  {F{\`e}ve}}, \bibinfo {author} {\bibfnamefont {A.}~\bibnamefont {Mah{\'e}}},
  \bibinfo {author} {\bibfnamefont {J.-M.}\ \bibnamefont {Berroir}}, \bibinfo
  {author} {\bibfnamefont {T.}~\bibnamefont {Kontos}}, \bibinfo {author}
  {\bibfnamefont {B.}~\bibnamefont {Pla{\c c}ais}}, \bibinfo {author}
  {\bibfnamefont {D.~C.}\ \bibnamefont {Glattli}}, \bibinfo {author}
  {\bibfnamefont {A.}~\bibnamefont {Cavanna}}, \bibinfo {author} {\bibfnamefont
  {B.}~\bibnamefont {Etienne}},\ and\ \bibinfo {author} {\bibfnamefont
  {Y.}~\bibnamefont {Jin}},\ }\href {https://doi.org/10.1126/science.1141243}
  {\bibfield  {journal} {\bibinfo  {journal} {Science}\ }\textbf {\bibinfo
  {volume} {316}},\ \bibinfo {pages} {1169} (\bibinfo {year}
  {2007})}\BibitemShut {NoStop}%
\bibitem [{\citenamefont {Bocquillon}\ \emph {et~al.}(2012)\citenamefont
  {Bocquillon}, \citenamefont {Parmentier}, \citenamefont {Grenier},
  \citenamefont {Berroir}, \citenamefont {Degiovanni}, \citenamefont {Glattli},
  \citenamefont {Pla\ifmmode~\mbox{\c{c}}\else \c{c}\fi{}ais}, \citenamefont
  {Cavanna}, \citenamefont {Jin},\ and\ \citenamefont
  {F\`eve}}]{bocquillon2012hbt}%
  \BibitemOpen
  \bibfield  {author} {\bibinfo {author} {\bibfnamefont {E.}~\bibnamefont
  {Bocquillon}}, \bibinfo {author} {\bibfnamefont {F.~D.}\ \bibnamefont
  {Parmentier}}, \bibinfo {author} {\bibfnamefont {C.}~\bibnamefont {Grenier}},
  \bibinfo {author} {\bibfnamefont {J.-M.}\ \bibnamefont {Berroir}}, \bibinfo
  {author} {\bibfnamefont {P.}~\bibnamefont {Degiovanni}}, \bibinfo {author}
  {\bibfnamefont {D.~C.}\ \bibnamefont {Glattli}}, \bibinfo {author}
  {\bibfnamefont {B.}~\bibnamefont {Pla\ifmmode~\mbox{\c{c}}\else
  \c{c}\fi{}ais}}, \bibinfo {author} {\bibfnamefont {A.}~\bibnamefont
  {Cavanna}}, \bibinfo {author} {\bibfnamefont {Y.}~\bibnamefont {Jin}},\ and\
  \bibinfo {author} {\bibfnamefont {G.}~\bibnamefont {F\`eve}},\ }\href
  {https://doi.org/10.1103/PhysRevLett.108.196803} {\bibfield  {journal}
  {\bibinfo  {journal} {Phys. Rev. Lett.}\ }\textbf {\bibinfo {volume} {108}},\
  \bibinfo {pages} {196803} (\bibinfo {year} {2012})}\BibitemShut {NoStop}%
\bibitem [{\citenamefont {Bocquillon}\ \emph {et~al.}(2013)\citenamefont
  {Bocquillon}, \citenamefont {Freulon}, \citenamefont {Berroir}, \citenamefont
  {Degiovanni}, \citenamefont {Pla{\c c}ais}, \citenamefont {Cavanna},
  \citenamefont {Jin},\ and\ \citenamefont
  {F{\`e}ve}}]{bocquillon2013coherenceInditinguishabilityIndependentSources}%
  \BibitemOpen
  \bibfield  {author} {\bibinfo {author} {\bibfnamefont {E.}~\bibnamefont
  {Bocquillon}}, \bibinfo {author} {\bibfnamefont {V.}~\bibnamefont {Freulon}},
  \bibinfo {author} {\bibfnamefont {J.-M.}\ \bibnamefont {Berroir}}, \bibinfo
  {author} {\bibfnamefont {P.}~\bibnamefont {Degiovanni}}, \bibinfo {author}
  {\bibfnamefont {B.}~\bibnamefont {Pla{\c c}ais}}, \bibinfo {author}
  {\bibfnamefont {A.}~\bibnamefont {Cavanna}}, \bibinfo {author} {\bibfnamefont
  {Y.}~\bibnamefont {Jin}},\ and\ \bibinfo {author} {\bibfnamefont
  {G.}~\bibnamefont {F{\`e}ve}},\ }\href
  {https://doi.org/10.1126/science.1232572} {\bibfield  {journal} {\bibinfo
  {journal} {Science}\ }\textbf {\bibinfo {volume} {339}},\ \bibinfo {pages}
  {1054} (\bibinfo {year} {2013})}\BibitemShut {NoStop}%
\bibitem [{\citenamefont {Dubois}\ \emph {et~al.}(2013)\citenamefont {Dubois},
  \citenamefont {Jullien}, \citenamefont {Portier}, \citenamefont {Roche},
  \citenamefont {Cavanna}, \citenamefont {Jin}, \citenamefont {Wegscheider},
  \citenamefont {Roulleau},\ and\ \citenamefont
  {Glattli}}]{dubois2013levitonEQO}%
  \BibitemOpen
  \bibfield  {author} {\bibinfo {author} {\bibfnamefont {J.}~\bibnamefont
  {Dubois}}, \bibinfo {author} {\bibfnamefont {T.}~\bibnamefont {Jullien}},
  \bibinfo {author} {\bibfnamefont {F.}~\bibnamefont {Portier}}, \bibinfo
  {author} {\bibfnamefont {P.}~\bibnamefont {Roche}}, \bibinfo {author}
  {\bibfnamefont {A.}~\bibnamefont {Cavanna}}, \bibinfo {author} {\bibfnamefont
  {Y.}~\bibnamefont {Jin}}, \bibinfo {author} {\bibfnamefont {W.}~\bibnamefont
  {Wegscheider}}, \bibinfo {author} {\bibfnamefont {P.}~\bibnamefont
  {Roulleau}},\ and\ \bibinfo {author} {\bibfnamefont {D.~C.}\ \bibnamefont
  {Glattli}},\ }\href {https://doi.org/10.1038/nature12713} {\bibfield
  {journal} {\bibinfo  {journal} {Nature}\ }\textbf {\bibinfo {volume} {502}},\
  \bibinfo {pages} {659} (\bibinfo {year} {2013})}\BibitemShut {NoStop}%
\bibitem [{\citenamefont {Blumenthal}\ \emph {et~al.}(2007)\citenamefont
  {Blumenthal}, \citenamefont {Kaestner}, \citenamefont {Li}, \citenamefont
  {Giblin}, \citenamefont {Janssen}, \citenamefont {Pepper}, \citenamefont
  {Anderson}, \citenamefont {Jones},\ and\ \citenamefont
  {Ritchie}}]{blumenthal2007GHzQuantChargePump}%
  \BibitemOpen
  \bibfield  {author} {\bibinfo {author} {\bibfnamefont {M.}~\bibnamefont
  {Blumenthal}}, \bibinfo {author} {\bibfnamefont {B.}~\bibnamefont
  {Kaestner}}, \bibinfo {author} {\bibfnamefont {L.}~\bibnamefont {Li}},
  \bibinfo {author} {\bibfnamefont {S.}~\bibnamefont {Giblin}}, \bibinfo
  {author} {\bibfnamefont {T.}~\bibnamefont {Janssen}}, \bibinfo {author}
  {\bibfnamefont {M.}~\bibnamefont {Pepper}}, \bibinfo {author} {\bibfnamefont
  {D.}~\bibnamefont {Anderson}}, \bibinfo {author} {\bibfnamefont
  {G.}~\bibnamefont {Jones}},\ and\ \bibinfo {author} {\bibfnamefont
  {D.}~\bibnamefont {Ritchie}},\ }\href {https://doi.org/10.1038/nphys582}
  {\bibfield  {journal} {\bibinfo  {journal} {Nat. Phys.}\ }\textbf {\bibinfo
  {volume} {3}},\ \bibinfo {pages} {343} (\bibinfo {year} {2007})}\BibitemShut
  {NoStop}%
\bibitem [{\citenamefont {Fletcher}\ \emph {et~al.}(2013)\citenamefont
  {Fletcher}, \citenamefont {See}, \citenamefont {Howe}, \citenamefont
  {Pepper}, \citenamefont {Giblin}, \citenamefont {Griffiths}, \citenamefont
  {Jones}, \citenamefont {Farrer}, \citenamefont {Ritchie}, \citenamefont
  {Janssen},\ and\ \citenamefont
  {Kataoka}}]{fletcher2013clockControlledEmission_SEwavePackets}%
  \BibitemOpen
  \bibfield  {author} {\bibinfo {author} {\bibfnamefont {J.~D.}\ \bibnamefont
  {Fletcher}}, \bibinfo {author} {\bibfnamefont {P.}~\bibnamefont {See}},
  \bibinfo {author} {\bibfnamefont {H.}~\bibnamefont {Howe}}, \bibinfo {author}
  {\bibfnamefont {M.}~\bibnamefont {Pepper}}, \bibinfo {author} {\bibfnamefont
  {S.~P.}\ \bibnamefont {Giblin}}, \bibinfo {author} {\bibfnamefont {J.~P.}\
  \bibnamefont {Griffiths}}, \bibinfo {author} {\bibfnamefont {G.~A.~C.}\
  \bibnamefont {Jones}}, \bibinfo {author} {\bibfnamefont {I.}~\bibnamefont
  {Farrer}}, \bibinfo {author} {\bibfnamefont {D.~A.}\ \bibnamefont {Ritchie}},
  \bibinfo {author} {\bibfnamefont {T.~J. B.~M.}\ \bibnamefont {Janssen}},\
  and\ \bibinfo {author} {\bibfnamefont {M.}~\bibnamefont {Kataoka}},\ }\href
  {https://doi.org/10.1103/PhysRevLett.111.216807} {\bibfield  {journal}
  {\bibinfo  {journal} {Phys. Rev. Lett.}\ }\textbf {\bibinfo {volume} {111}},\
  \bibinfo {pages} {216807} (\bibinfo {year} {2013})}\BibitemShut {NoStop}%
\bibitem [{\citenamefont {Ubbelohde}\ \emph {et~al.}(2015)\citenamefont
  {Ubbelohde}, \citenamefont {Hohls}, \citenamefont {Kashcheyevs},
  \citenamefont {Wagner}, \citenamefont {Fricke}, \citenamefont {K{\"a}stner},
  \citenamefont {Pierz}, \citenamefont {Schumacher},\ and\ \citenamefont
  {Haug}}]{ubbelohde2015partitioningOnDemandElecPairs}%
  \BibitemOpen
  \bibfield  {author} {\bibinfo {author} {\bibfnamefont {N.}~\bibnamefont
  {Ubbelohde}}, \bibinfo {author} {\bibfnamefont {F.}~\bibnamefont {Hohls}},
  \bibinfo {author} {\bibfnamefont {V.}~\bibnamefont {Kashcheyevs}}, \bibinfo
  {author} {\bibfnamefont {T.}~\bibnamefont {Wagner}}, \bibinfo {author}
  {\bibfnamefont {L.}~\bibnamefont {Fricke}}, \bibinfo {author} {\bibfnamefont
  {B.}~\bibnamefont {K{\"a}stner}}, \bibinfo {author} {\bibfnamefont
  {K.}~\bibnamefont {Pierz}}, \bibinfo {author} {\bibfnamefont {H.~W.}\
  \bibnamefont {Schumacher}},\ and\ \bibinfo {author} {\bibfnamefont {R.~J.}\
  \bibnamefont {Haug}},\ }\href {https://doi.org/10.1038/nnano.2014.275}
  {\bibfield  {journal} {\bibinfo  {journal} {Nat. Nanotechnol.}\ }\textbf
  {\bibinfo {volume} {10}},\ \bibinfo {pages} {46} (\bibinfo {year}
  {2015})}\BibitemShut {NoStop}%
\bibitem [{\citenamefont {Hermelin}\ \emph {et~al.}(2011)\citenamefont
  {Hermelin}, \citenamefont {Takada}, \citenamefont {Yamamoto}, \citenamefont
  {Tarucha}, \citenamefont {Wieck}, \citenamefont {Saminadayar}, \citenamefont
  {B{\"a}uerle},\ and\ \citenamefont {Meunier}}]{hermelin2011SawEqo}%
  \BibitemOpen
  \bibfield  {author} {\bibinfo {author} {\bibfnamefont {S.}~\bibnamefont
  {Hermelin}}, \bibinfo {author} {\bibfnamefont {S.}~\bibnamefont {Takada}},
  \bibinfo {author} {\bibfnamefont {M.}~\bibnamefont {Yamamoto}}, \bibinfo
  {author} {\bibfnamefont {S.}~\bibnamefont {Tarucha}}, \bibinfo {author}
  {\bibfnamefont {A.~D.}\ \bibnamefont {Wieck}}, \bibinfo {author}
  {\bibfnamefont {L.}~\bibnamefont {Saminadayar}}, \bibinfo {author}
  {\bibfnamefont {C.}~\bibnamefont {B{\"a}uerle}},\ and\ \bibinfo {author}
  {\bibfnamefont {T.}~\bibnamefont {Meunier}},\ }\href
  {https://doi.org/10.1038/nature10416} {\bibfield  {journal} {\bibinfo
  {journal} {Nature}\ }\textbf {\bibinfo {volume} {477}},\ \bibinfo {pages}
  {435} (\bibinfo {year} {2011})}\BibitemShut {NoStop}%
\bibitem [{\citenamefont {McNeil}\ \emph {et~al.}(2011)\citenamefont {McNeil},
  \citenamefont {Kataoka}, \citenamefont {Ford}, \citenamefont {Barnes},
  \citenamefont {Anderson}, \citenamefont {Jones}, \citenamefont {Farrer},\
  and\ \citenamefont {Ritchie}}]{mcneil2011SingleElTransferDistantQDs}%
  \BibitemOpen
  \bibfield  {author} {\bibinfo {author} {\bibfnamefont {R.~P.~G.}\
  \bibnamefont {McNeil}}, \bibinfo {author} {\bibfnamefont {M.}~\bibnamefont
  {Kataoka}}, \bibinfo {author} {\bibfnamefont {C.~J.~B.}\ \bibnamefont
  {Ford}}, \bibinfo {author} {\bibfnamefont {C.~H.~W.}\ \bibnamefont {Barnes}},
  \bibinfo {author} {\bibfnamefont {D.}~\bibnamefont {Anderson}}, \bibinfo
  {author} {\bibfnamefont {G.~A.~C.}\ \bibnamefont {Jones}}, \bibinfo {author}
  {\bibfnamefont {I.}~\bibnamefont {Farrer}},\ and\ \bibinfo {author}
  {\bibfnamefont {D.~A.}\ \bibnamefont {Ritchie}},\ }\href
  {https://doi.org/10.1038/nature10444} {\bibfield  {journal} {\bibinfo
  {journal} {Nature}\ }\textbf {\bibinfo {volume} {477}},\ \bibinfo {pages}
  {439} (\bibinfo {year} {2011})}\BibitemShut {NoStop}%
\bibitem [{\citenamefont {Hadfield}(2009)}]{hadfield2009singlePhotonDetectors}%
  \BibitemOpen
  \bibfield  {author} {\bibinfo {author} {\bibfnamefont {R.~H.}\ \bibnamefont
  {Hadfield}},\ }\href {https://doi.org/10.1038/nphoton.2009.230} {\bibfield
  {journal} {\bibinfo  {journal} {Nat. Photonics}\ }\textbf {\bibinfo {volume}
  {3}},\ \bibinfo {pages} {696} (\bibinfo {year} {2009})}\BibitemShut {NoStop}%
\bibitem [{sup()}]{supp}%
  \BibitemOpen
  \href@noop {} {}\bibinfo {note} {See Supplemental Material at
  [UrlInsertedByPublisher] for additional information on sample B, sample
  operation and characterization.}\BibitemShut {Stop}%
\bibitem [{\citenamefont {Gerster}\ \emph {et~al.}(2019)\citenamefont
  {Gerster}, \citenamefont {Mueller}, \citenamefont {Freise}, \citenamefont
  {Reifert}, \citenamefont {Maradan}, \citenamefont {Hinze}, \citenamefont
  {Weimann}, \citenamefont {Marx}, \citenamefont {Pierz}, \citenamefont
  {Schumacher}, \citenamefont {Hohls},\ and\ \citenamefont
  {Ubbelohde}}]{gerster2018qd}%
  \BibitemOpen
  \bibfield  {author} {\bibinfo {author} {\bibfnamefont {T.}~\bibnamefont
  {Gerster}}, \bibinfo {author} {\bibfnamefont {A.}~\bibnamefont {Mueller}},
  \bibinfo {author} {\bibfnamefont {L.}~\bibnamefont {Freise}}, \bibinfo
  {author} {\bibfnamefont {D.}~\bibnamefont {Reifert}}, \bibinfo {author}
  {\bibfnamefont {D.}~\bibnamefont {Maradan}}, \bibinfo {author} {\bibfnamefont
  {P.}~\bibnamefont {Hinze}}, \bibinfo {author} {\bibfnamefont
  {T.}~\bibnamefont {Weimann}}, \bibinfo {author} {\bibfnamefont
  {H.}~\bibnamefont {Marx}}, \bibinfo {author} {\bibfnamefont {K.}~\bibnamefont
  {Pierz}}, \bibinfo {author} {\bibfnamefont {H.~W.}\ \bibnamefont
  {Schumacher}}, \bibinfo {author} {\bibfnamefont {F.}~\bibnamefont {Hohls}},\
  and\ \bibinfo {author} {\bibfnamefont {N.}~\bibnamefont {Ubbelohde}},\ }\href
  {https://doi.org/10.1088/1681-7575/aaf4aa} {\bibfield  {journal} {\bibinfo
  {journal} {Metrologia}\ }\textbf {\bibinfo {volume} {56}},\ \bibinfo {pages}
  {014002} (\bibinfo {year} {2019})}\BibitemShut {NoStop}%
\bibitem [{\citenamefont {Coleridge}\ \emph {et~al.}(1994)\citenamefont
  {Coleridge}, \citenamefont {Zawadzki},\ and\ \citenamefont
  {Sachrajda}}]{coleridge1994resistivityPeakValues}%
  \BibitemOpen
  \bibfield  {author} {\bibinfo {author} {\bibfnamefont {P.~T.}\ \bibnamefont
  {Coleridge}}, \bibinfo {author} {\bibfnamefont {P.}~\bibnamefont
  {Zawadzki}},\ and\ \bibinfo {author} {\bibfnamefont {A.~S.}\ \bibnamefont
  {Sachrajda}},\ }\href {https://doi.org/10.1103/PhysRevB.49.10798} {\bibfield
  {journal} {\bibinfo  {journal} {Phys. Rev. B}\ }\textbf {\bibinfo {volume}
  {49}},\ \bibinfo {pages} {10798} (\bibinfo {year} {1994})}\BibitemShut
  {NoStop}%
\bibitem [{\citenamefont {Kaestner}\ \emph {et~al.}(2008)\citenamefont
  {Kaestner}, \citenamefont {Kashcheyevs}, \citenamefont {Amakawa},
  \citenamefont {Blumenthal}, \citenamefont {Li}, \citenamefont {Janssen},
  \citenamefont {Hein}, \citenamefont {Pierz}, \citenamefont {Weimann},
  \citenamefont {Siegner},\ and\ \citenamefont
  {Schumacher}}]{kaestner2008singleParamNonadQuantChargePump}%
  \BibitemOpen
  \bibfield  {author} {\bibinfo {author} {\bibfnamefont {B.}~\bibnamefont
  {Kaestner}}, \bibinfo {author} {\bibfnamefont {V.}~\bibnamefont
  {Kashcheyevs}}, \bibinfo {author} {\bibfnamefont {S.}~\bibnamefont
  {Amakawa}}, \bibinfo {author} {\bibfnamefont {M.~D.}\ \bibnamefont
  {Blumenthal}}, \bibinfo {author} {\bibfnamefont {L.}~\bibnamefont {Li}},
  \bibinfo {author} {\bibfnamefont {T.~J. B.~M.}\ \bibnamefont {Janssen}},
  \bibinfo {author} {\bibfnamefont {G.}~\bibnamefont {Hein}}, \bibinfo {author}
  {\bibfnamefont {K.}~\bibnamefont {Pierz}}, \bibinfo {author} {\bibfnamefont
  {T.}~\bibnamefont {Weimann}}, \bibinfo {author} {\bibfnamefont
  {U.}~\bibnamefont {Siegner}},\ and\ \bibinfo {author} {\bibfnamefont {H.~W.}\
  \bibnamefont {Schumacher}},\ }\href
  {https://doi.org/10.1103/PhysRevB.77.153301} {\bibfield  {journal} {\bibinfo
  {journal} {Phys. Rev. B}\ }\textbf {\bibinfo {volume} {77}},\ \bibinfo
  {pages} {153301} (\bibinfo {year} {2008})}\BibitemShut {NoStop}%
\bibitem [{\citenamefont {Maire}\ \emph {et~al.}(2008)\citenamefont {Maire},
  \citenamefont {Hohls}, \citenamefont {Kaestner}, \citenamefont {Pierz},
  \citenamefont {Schumacher},\ and\ \citenamefont
  {Haug}}]{maire2008noiseMeasSEP}%
  \BibitemOpen
  \bibfield  {author} {\bibinfo {author} {\bibfnamefont {N.}~\bibnamefont
  {Maire}}, \bibinfo {author} {\bibfnamefont {F.}~\bibnamefont {Hohls}},
  \bibinfo {author} {\bibfnamefont {B.}~\bibnamefont {Kaestner}}, \bibinfo
  {author} {\bibfnamefont {K.}~\bibnamefont {Pierz}}, \bibinfo {author}
  {\bibfnamefont {H.~W.}\ \bibnamefont {Schumacher}},\ and\ \bibinfo {author}
  {\bibfnamefont {R.~J.}\ \bibnamefont {Haug}},\ }\href
  {https://doi.org/10.1063/1.2885076} {\bibfield  {journal} {\bibinfo
  {journal} {Appl. Phys. Lett.}\ }\textbf {\bibinfo {volume} {92}},\ \bibinfo
  {pages} {082112} (\bibinfo {year} {2008})}\BibitemShut {NoStop}%
\bibitem [{\citenamefont {Leicht}\ \emph {et~al.}(2011)\citenamefont {Leicht},
  \citenamefont {Mirovsky}, \citenamefont {Kaestner}, \citenamefont {Hohls},
  \citenamefont {Kashcheyevs}, \citenamefont {Kurganova}, \citenamefont
  {Zeitler}, \citenamefont {Weimann}, \citenamefont {Pierz},\ and\
  \citenamefont {Schumacher}}]{leicht2011SEPgenInQHedgestates}%
  \BibitemOpen
  \bibfield  {author} {\bibinfo {author} {\bibfnamefont {C.}~\bibnamefont
  {Leicht}}, \bibinfo {author} {\bibfnamefont {P.}~\bibnamefont {Mirovsky}},
  \bibinfo {author} {\bibfnamefont {B.}~\bibnamefont {Kaestner}}, \bibinfo
  {author} {\bibfnamefont {F.}~\bibnamefont {Hohls}}, \bibinfo {author}
  {\bibfnamefont {V.}~\bibnamefont {Kashcheyevs}}, \bibinfo {author}
  {\bibfnamefont {E.~V.}\ \bibnamefont {Kurganova}}, \bibinfo {author}
  {\bibfnamefont {U.}~\bibnamefont {Zeitler}}, \bibinfo {author} {\bibfnamefont
  {T.}~\bibnamefont {Weimann}}, \bibinfo {author} {\bibfnamefont
  {K.}~\bibnamefont {Pierz}},\ and\ \bibinfo {author} {\bibfnamefont {H.~W.}\
  \bibnamefont {Schumacher}},\ }\href
  {https://doi.org/10.1088/0268-1242/26/5/055010} {\bibfield  {journal}
  {\bibinfo  {journal} {Semicond. Sci. Technol.}\ }\textbf {\bibinfo {volume}
  {26}},\ \bibinfo {pages} {055010} (\bibinfo {year} {2011})}\BibitemShut
  {NoStop}%
\bibitem [{\citenamefont {Stein}\ \emph {et~al.}(2015)\citenamefont {Stein},
  \citenamefont {Drung}, \citenamefont {Fricke}, \citenamefont {Scherer},
  \citenamefont {Hohls}, \citenamefont {Leicht}, \citenamefont {G{\"o}tz},
  \citenamefont {Krause}, \citenamefont {Behr}, \citenamefont {Pesel},
  \citenamefont {Pierz}, \citenamefont {Siegner}, \citenamefont {Ahlers},\ and\
  \citenamefont {Schumacher}}]{stein2015subppm}%
  \BibitemOpen
  \bibfield  {author} {\bibinfo {author} {\bibfnamefont {F.}~\bibnamefont
  {Stein}}, \bibinfo {author} {\bibfnamefont {D.}~\bibnamefont {Drung}},
  \bibinfo {author} {\bibfnamefont {L.}~\bibnamefont {Fricke}}, \bibinfo
  {author} {\bibfnamefont {H.}~\bibnamefont {Scherer}}, \bibinfo {author}
  {\bibfnamefont {F.}~\bibnamefont {Hohls}}, \bibinfo {author} {\bibfnamefont
  {C.}~\bibnamefont {Leicht}}, \bibinfo {author} {\bibfnamefont
  {M.}~\bibnamefont {G{\"o}tz}}, \bibinfo {author} {\bibfnamefont
  {C.}~\bibnamefont {Krause}}, \bibinfo {author} {\bibfnamefont
  {R.}~\bibnamefont {Behr}}, \bibinfo {author} {\bibfnamefont {E.}~\bibnamefont
  {Pesel}}, \bibinfo {author} {\bibfnamefont {K.}~\bibnamefont {Pierz}},
  \bibinfo {author} {\bibfnamefont {U.}~\bibnamefont {Siegner}}, \bibinfo
  {author} {\bibfnamefont {F.~J.}\ \bibnamefont {Ahlers}},\ and\ \bibinfo
  {author} {\bibfnamefont {H.~W.}\ \bibnamefont {Schumacher}},\ }\href
  {https://doi.org/10.1063/1.4930142} {\bibfield  {journal} {\bibinfo
  {journal} {Appl. Phys. Lett.}\ }\textbf {\bibinfo {volume} {107}},\ \bibinfo
  {pages} {103501} (\bibinfo {year} {2015})}\BibitemShut {NoStop}%
\bibitem [{\citenamefont {Tewari}\ \emph {et~al.}(2016)\citenamefont {Tewari},
  \citenamefont {Roulleau}, \citenamefont {Grenier}, \citenamefont {Portier},
  \citenamefont {Cavanna}, \citenamefont {Gennser}, \citenamefont {Mailly},\
  and\ \citenamefont {Roche}}]{tewari2016quantumCoherenceAboreEF}%
  \BibitemOpen
  \bibfield  {author} {\bibinfo {author} {\bibfnamefont {S.}~\bibnamefont
  {Tewari}}, \bibinfo {author} {\bibfnamefont {P.}~\bibnamefont {Roulleau}},
  \bibinfo {author} {\bibfnamefont {C.}~\bibnamefont {Grenier}}, \bibinfo
  {author} {\bibfnamefont {F.}~\bibnamefont {Portier}}, \bibinfo {author}
  {\bibfnamefont {A.}~\bibnamefont {Cavanna}}, \bibinfo {author} {\bibfnamefont
  {U.}~\bibnamefont {Gennser}}, \bibinfo {author} {\bibfnamefont
  {D.}~\bibnamefont {Mailly}},\ and\ \bibinfo {author} {\bibfnamefont
  {P.}~\bibnamefont {Roche}},\ }\href
  {https://doi.org/10.1103/PhysRevB.93.035420} {\bibfield  {journal} {\bibinfo
  {journal} {Phys. Rev. B}\ }\textbf {\bibinfo {volume} {93}},\ \bibinfo
  {pages} {035420} (\bibinfo {year} {2016})}\BibitemShut {NoStop}%
\bibitem [{\citenamefont {Nigg}\ and\ \citenamefont
  {Lunde}(2016)}]{nigg2016decoherenceHighEnElQHE}%
  \BibitemOpen
  \bibfield  {author} {\bibinfo {author} {\bibfnamefont {S.~E.}\ \bibnamefont
  {Nigg}}\ and\ \bibinfo {author} {\bibfnamefont {A.~M.}\ \bibnamefont
  {Lunde}},\ }\href {https://doi.org/10.1103/PhysRevB.94.041407} {\bibfield
  {journal} {\bibinfo  {journal} {Phys. Rev. B}\ }\textbf {\bibinfo {volume}
  {94}},\ \bibinfo {pages} {041407(R)} (\bibinfo {year} {2016})}\BibitemShut
  {NoStop}%
\bibitem [{\citenamefont {Bocquillon}\ \emph {et~al.}(2014)\citenamefont
  {Bocquillon}, \citenamefont {Freulon}, \citenamefont {Parmentier},
  \citenamefont {Berroir}, \citenamefont {Plaçais}, \citenamefont {Wahl},
  \citenamefont {Rech}, \citenamefont {Jonckheere}, \citenamefont {Martin},
  \citenamefont {Grenier}, \citenamefont {Ferraro}, \citenamefont
  {Degiovanni},\ and\ \citenamefont
  {Fève}}]{bocquillon2013review_EQOinBallisticConductors}%
  \BibitemOpen
  \bibfield  {author} {\bibinfo {author} {\bibfnamefont {E.}~\bibnamefont
  {Bocquillon}}, \bibinfo {author} {\bibfnamefont {V.}~\bibnamefont {Freulon}},
  \bibinfo {author} {\bibfnamefont {F.~D.}\ \bibnamefont {Parmentier}},
  \bibinfo {author} {\bibfnamefont {J.-M.}\ \bibnamefont {Berroir}}, \bibinfo
  {author} {\bibfnamefont {B.}~\bibnamefont {Plaçais}}, \bibinfo {author}
  {\bibfnamefont {C.}~\bibnamefont {Wahl}}, \bibinfo {author} {\bibfnamefont
  {J.}~\bibnamefont {Rech}}, \bibinfo {author} {\bibfnamefont {T.}~\bibnamefont
  {Jonckheere}}, \bibinfo {author} {\bibfnamefont {T.}~\bibnamefont {Martin}},
  \bibinfo {author} {\bibfnamefont {C.}~\bibnamefont {Grenier}}, \bibinfo
  {author} {\bibfnamefont {D.}~\bibnamefont {Ferraro}}, \bibinfo {author}
  {\bibfnamefont {P.}~\bibnamefont {Degiovanni}},\ and\ \bibinfo {author}
  {\bibfnamefont {G.}~\bibnamefont {Fève}},\ }\href
  {https://doi.org/10.1002/andp.201300181} {\bibfield  {journal} {\bibinfo
  {journal} {Ann. Phys.}\ }\textbf {\bibinfo {volume} {526}},\ \bibinfo {pages}
  {1} (\bibinfo {year} {2014})}\BibitemShut {NoStop}%
\bibitem [{\citenamefont {Krähenmann}\ \emph {et~al.}(2019)\citenamefont
  {Krähenmann}, \citenamefont {Fischer}, \citenamefont {Röösli},
  \citenamefont {Ihn}, \citenamefont {Reichl}, \citenamefont {Wegscheider},
  \citenamefont {Ensslin}, \citenamefont {Gefen},\ and\ \citenamefont
  {Meir}}]{kraehenmann2019augerSpectrSE_QHE}%
  \BibitemOpen
  \bibfield  {author} {\bibinfo {author} {\bibfnamefont {T.}~\bibnamefont
  {Krähenmann}}, \bibinfo {author} {\bibfnamefont {S.~G.}\ \bibnamefont
  {Fischer}}, \bibinfo {author} {\bibfnamefont {M.}~\bibnamefont {Röösli}},
  \bibinfo {author} {\bibfnamefont {T.}~\bibnamefont {Ihn}}, \bibinfo {author}
  {\bibfnamefont {C.}~\bibnamefont {Reichl}}, \bibinfo {author} {\bibfnamefont
  {W.}~\bibnamefont {Wegscheider}}, \bibinfo {author} {\bibfnamefont
  {K.}~\bibnamefont {Ensslin}}, \bibinfo {author} {\bibfnamefont
  {Y.}~\bibnamefont {Gefen}},\ and\ \bibinfo {author} {\bibfnamefont
  {Y.}~\bibnamefont {Meir}},\ }\href
  {https://doi.org/10.1038/s41467-019-11888-1} {\bibfield  {journal} {\bibinfo
  {journal} {Nat. Commun.}\ }\textbf {\bibinfo {volume} {10}},\ \bibinfo
  {pages} {3915} (\bibinfo {year} {2019})}\BibitemShut {NoStop}%
\bibitem [{\citenamefont {Rodriguez}\ \emph {et~al.}()\citenamefont
  {Rodriguez}, \citenamefont {Parmentier}, \citenamefont {Roulleau},
  \citenamefont {Gennser}, \citenamefont {Cavanna}, \citenamefont {Portier},
  \citenamefont {Mailly},\ and\ \citenamefont
  {Roche}}]{rodriguez2019stronEnergyRelaxQHE}%
  \BibitemOpen
  \bibfield  {author} {\bibinfo {author} {\bibfnamefont {R.~H.}\ \bibnamefont
  {Rodriguez}}, \bibinfo {author} {\bibfnamefont {F.~D.}\ \bibnamefont
  {Parmentier}}, \bibinfo {author} {\bibfnamefont {P.}~\bibnamefont
  {Roulleau}}, \bibinfo {author} {\bibfnamefont {U.}~\bibnamefont {Gennser}},
  \bibinfo {author} {\bibfnamefont {A.}~\bibnamefont {Cavanna}}, \bibinfo
  {author} {\bibfnamefont {F.}~\bibnamefont {Portier}}, \bibinfo {author}
  {\bibfnamefont {D.}~\bibnamefont {Mailly}},\ and\ \bibinfo {author}
  {\bibfnamefont {P.}~\bibnamefont {Roche}},\ }\href@noop {} {\ }\Eprint
  {https://arxiv.org/abs/1903.05919} {arXiv:1903.05919} \BibitemShut {NoStop}%
\bibitem [{\citenamefont {Kataoka}\ \emph {et~al.}(2016)\citenamefont
  {Kataoka}, \citenamefont {Johnson}, \citenamefont {Emary}, \citenamefont
  {See}, \citenamefont {Griffiths}, \citenamefont {Jones}, \citenamefont
  {Farrer}, \citenamefont {Ritchie}, \citenamefont {Pepper},\ and\
  \citenamefont {Janssen}}]{kataoka2016TOF_SEwavepacket_edgeSate}%
  \BibitemOpen
  \bibfield  {author} {\bibinfo {author} {\bibfnamefont {M.}~\bibnamefont
  {Kataoka}}, \bibinfo {author} {\bibfnamefont {N.}~\bibnamefont {Johnson}},
  \bibinfo {author} {\bibfnamefont {C.}~\bibnamefont {Emary}}, \bibinfo
  {author} {\bibfnamefont {P.}~\bibnamefont {See}}, \bibinfo {author}
  {\bibfnamefont {J.~P.}\ \bibnamefont {Griffiths}}, \bibinfo {author}
  {\bibfnamefont {G.~A.~C.}\ \bibnamefont {Jones}}, \bibinfo {author}
  {\bibfnamefont {I.}~\bibnamefont {Farrer}}, \bibinfo {author} {\bibfnamefont
  {D.~A.}\ \bibnamefont {Ritchie}}, \bibinfo {author} {\bibfnamefont
  {M.}~\bibnamefont {Pepper}},\ and\ \bibinfo {author} {\bibfnamefont {T.~J.
  B.~M.}\ \bibnamefont {Janssen}},\ }\href
  {https://doi.org/10.1103/PhysRevLett.116.126803} {\bibfield  {journal}
  {\bibinfo  {journal} {Phys. Rev. Lett.}\ }\textbf {\bibinfo {volume} {116}},\
  \bibinfo {pages} {126803} (\bibinfo {year} {2016})}\BibitemShut {NoStop}%
\bibitem [{\citenamefont {Johnson}\ \emph {et~al.}(2018)\citenamefont
  {Johnson}, \citenamefont {Emary}, \citenamefont {Ryu}, \citenamefont {Sim},
  \citenamefont {See}, \citenamefont {Fletcher}, \citenamefont {Griffiths},
  \citenamefont {Jones}, \citenamefont {Farrer}, \citenamefont {Ritchie},
  \citenamefont {Pepper}, \citenamefont {Janssen},\ and\ \citenamefont
  {Kataoka}}]{johnson2017phononEmissionRate}%
  \BibitemOpen
  \bibfield  {author} {\bibinfo {author} {\bibfnamefont {N.}~\bibnamefont
  {Johnson}}, \bibinfo {author} {\bibfnamefont {C.}~\bibnamefont {Emary}},
  \bibinfo {author} {\bibfnamefont {S.}~\bibnamefont {Ryu}}, \bibinfo {author}
  {\bibfnamefont {H.-S.}\ \bibnamefont {Sim}}, \bibinfo {author} {\bibfnamefont
  {P.}~\bibnamefont {See}}, \bibinfo {author} {\bibfnamefont {J.~D.}\
  \bibnamefont {Fletcher}}, \bibinfo {author} {\bibfnamefont {J.~P.}\
  \bibnamefont {Griffiths}}, \bibinfo {author} {\bibfnamefont {G.~A.~C.}\
  \bibnamefont {Jones}}, \bibinfo {author} {\bibfnamefont {I.}~\bibnamefont
  {Farrer}}, \bibinfo {author} {\bibfnamefont {D.~A.}\ \bibnamefont {Ritchie}},
  \bibinfo {author} {\bibfnamefont {M.}~\bibnamefont {Pepper}}, \bibinfo
  {author} {\bibfnamefont {T.~J. B.~M.}\ \bibnamefont {Janssen}},\ and\
  \bibinfo {author} {\bibfnamefont {M.}~\bibnamefont {Kataoka}},\ }\href
  {https://doi.org/10.1103/PhysRevLett.121.137703} {\bibfield  {journal}
  {\bibinfo  {journal} {Phys. Rev. Lett.}\ }\textbf {\bibinfo {volume} {121}},\
  \bibinfo {pages} {137703} (\bibinfo {year} {2018})}\BibitemShut {NoStop}%
\bibitem [{\citenamefont {Palevski}\ \emph {et~al.}(1989)\citenamefont
  {Palevski}, \citenamefont {Heiblum}, \citenamefont {Umbach}, \citenamefont
  {Knoedler}, \citenamefont {Broers},\ and\ \citenamefont
  {Koch}}]{palevski1989latTunneling+BallisTransp+Spectrosc2DEG}%
  \BibitemOpen
  \bibfield  {author} {\bibinfo {author} {\bibfnamefont {A.}~\bibnamefont
  {Palevski}}, \bibinfo {author} {\bibfnamefont {M.}~\bibnamefont {Heiblum}},
  \bibinfo {author} {\bibfnamefont {C.~P.}\ \bibnamefont {Umbach}}, \bibinfo
  {author} {\bibfnamefont {C.~M.}\ \bibnamefont {Knoedler}}, \bibinfo {author}
  {\bibfnamefont {A.~N.}\ \bibnamefont {Broers}},\ and\ \bibinfo {author}
  {\bibfnamefont {R.~H.}\ \bibnamefont {Koch}},\ }\href
  {https://doi.org/10.1103/PhysRevLett.62.1776} {\bibfield  {journal} {\bibinfo
   {journal} {Phys. Rev. Lett.}\ }\textbf {\bibinfo {volume} {62}},\ \bibinfo
  {pages} {1776} (\bibinfo {year} {1989})}\BibitemShut {NoStop}%
\bibitem [{\citenamefont {Ota}\ \emph {et~al.}(2019)\citenamefont {Ota},
  \citenamefont {Akiyama}, \citenamefont {Hashisaka}, \citenamefont {Muraki},\
  and\ \citenamefont {Fujisawa}}]{ota2019spectrosHotEl}%
  \BibitemOpen
  \bibfield  {author} {\bibinfo {author} {\bibfnamefont {T.}~\bibnamefont
  {Ota}}, \bibinfo {author} {\bibfnamefont {S.}~\bibnamefont {Akiyama}},
  \bibinfo {author} {\bibfnamefont {M.}~\bibnamefont {Hashisaka}}, \bibinfo
  {author} {\bibfnamefont {K.}~\bibnamefont {Muraki}},\ and\ \bibinfo {author}
  {\bibfnamefont {T.}~\bibnamefont {Fujisawa}},\ }\href
  {https://doi.org/10.1103/PhysRevB.99.085310} {\bibfield  {journal} {\bibinfo
  {journal} {Phys. Rev. B}\ }\textbf {\bibinfo {volume} {99}},\ \bibinfo
  {pages} {085310} (\bibinfo {year} {2019})}\BibitemShut {NoStop}%
\bibitem [{fn6()}]{fn6}%
  \BibitemOpen
  \href@noop {} {}\bibinfo {note} {The different geometry of the waveguide
  section in sample B yields lower values of $\lambda_\mathrm{WG}$ compared to
  sample A.}\BibitemShut {Stop}%
\bibitem [{\citenamefont {Taubert}\ \emph {et~al.}(2011)\citenamefont
  {Taubert}, \citenamefont {Tomaras}, \citenamefont {Schinner}, \citenamefont
  {Tranitz}, \citenamefont {Wegscheider}, \citenamefont {Kehrein},\ and\
  \citenamefont {Ludwig}}]{taubert2011phononRelaxation}%
  \BibitemOpen
  \bibfield  {author} {\bibinfo {author} {\bibfnamefont {D.}~\bibnamefont
  {Taubert}}, \bibinfo {author} {\bibfnamefont {C.}~\bibnamefont {Tomaras}},
  \bibinfo {author} {\bibfnamefont {G.~J.}\ \bibnamefont {Schinner}}, \bibinfo
  {author} {\bibfnamefont {H.~P.}\ \bibnamefont {Tranitz}}, \bibinfo {author}
  {\bibfnamefont {W.}~\bibnamefont {Wegscheider}}, \bibinfo {author}
  {\bibfnamefont {S.}~\bibnamefont {Kehrein}},\ and\ \bibinfo {author}
  {\bibfnamefont {S.}~\bibnamefont {Ludwig}},\ }\href
  {https://doi.org/10.1103/PhysRevB.83.235404} {\bibfield  {journal} {\bibinfo
  {journal} {Phys. Rev. B}\ }\textbf {\bibinfo {volume} {83}},\ \bibinfo
  {pages} {235404} (\bibinfo {year} {2011})}\BibitemShut {NoStop}%
\bibitem [{\citenamefont {Emary}\ \emph {et~al.}(2016)\citenamefont {Emary},
  \citenamefont {Dyson}, \citenamefont {Ryu}, \citenamefont {Sim},\ and\
  \citenamefont {Kataoka}}]{emary2016PhononEmission+ATD_SEP}%
  \BibitemOpen
  \bibfield  {author} {\bibinfo {author} {\bibfnamefont {C.}~\bibnamefont
  {Emary}}, \bibinfo {author} {\bibfnamefont {A.}~\bibnamefont {Dyson}},
  \bibinfo {author} {\bibfnamefont {S.}~\bibnamefont {Ryu}}, \bibinfo {author}
  {\bibfnamefont {H.-S.}\ \bibnamefont {Sim}},\ and\ \bibinfo {author}
  {\bibfnamefont {M.}~\bibnamefont {Kataoka}},\ }\href
  {https://doi.org/10.1103/PhysRevB.93.035436} {\bibfield  {journal} {\bibinfo
  {journal} {Phys. Rev. B}\ }\textbf {\bibinfo {volume} {93}},\ \bibinfo
  {pages} {035436} (\bibinfo {year} {2016})}\BibitemShut {NoStop}%
\bibitem [{\citenamefont {Emary}\ \emph {et~al.}(2019)\citenamefont {Emary},
  \citenamefont {Clark}, \citenamefont {Kataoka},\ and\ \citenamefont
  {Johnson}}]{emary2018phonon}%
  \BibitemOpen
  \bibfield  {author} {\bibinfo {author} {\bibfnamefont {C.}~\bibnamefont
  {Emary}}, \bibinfo {author} {\bibfnamefont {L.~A.}\ \bibnamefont {Clark}},
  \bibinfo {author} {\bibfnamefont {M.}~\bibnamefont {Kataoka}},\ and\ \bibinfo
  {author} {\bibfnamefont {N.}~\bibnamefont {Johnson}},\ }\href
  {https://doi.org/10.1103/PhysRevB.99.045306} {\bibfield  {journal} {\bibinfo
  {journal} {Phys. Rev. B}\ }\textbf {\bibinfo {volume} {99}},\ \bibinfo
  {pages} {045306} (\bibinfo {year} {2019})}\BibitemShut {NoStop}%
\bibitem [{fn5()}]{fn5}%
  \BibitemOpen
  \href@noop {} {}\bibinfo {note} {Comparison bases on similar experimental
  conditions, i.e., matching magnitude of electron velocities, here inferred
  from magnetic-field dependence of $\lambda_\mathrm{WG}$
  \cite{ota2019spectrosHotEl}.}\BibitemShut {Stop}%
\bibitem [{\citenamefont {Waldie}\ \emph {et~al.}(2015)\citenamefont {Waldie},
  \citenamefont {See}, \citenamefont {Kashcheyevs}, \citenamefont {Griffiths},
  \citenamefont {Farrer}, \citenamefont {Jones}, \citenamefont {Ritchie},
  \citenamefont {Janssen},\ and\ \citenamefont
  {Kataoka}}]{waldie2015elWavePacketsSEP}%
  \BibitemOpen
  \bibfield  {author} {\bibinfo {author} {\bibfnamefont {J.}~\bibnamefont
  {Waldie}}, \bibinfo {author} {\bibfnamefont {P.}~\bibnamefont {See}},
  \bibinfo {author} {\bibfnamefont {V.}~\bibnamefont {Kashcheyevs}}, \bibinfo
  {author} {\bibfnamefont {J.~P.}\ \bibnamefont {Griffiths}}, \bibinfo {author}
  {\bibfnamefont {I.}~\bibnamefont {Farrer}}, \bibinfo {author} {\bibfnamefont
  {G.~A.~C.}\ \bibnamefont {Jones}}, \bibinfo {author} {\bibfnamefont {D.~A.}\
  \bibnamefont {Ritchie}}, \bibinfo {author} {\bibfnamefont {T.~J. B.~M.}\
  \bibnamefont {Janssen}},\ and\ \bibinfo {author} {\bibfnamefont
  {M.}~\bibnamefont {Kataoka}},\ }\href
  {https://doi.org/10.1103/PhysRevB.92.125305} {\bibfield  {journal} {\bibinfo
  {journal} {Phys. Rev. B}\ }\textbf {\bibinfo {volume} {92}},\ \bibinfo
  {pages} {125305} (\bibinfo {year} {2015})}\BibitemShut {NoStop}%
\bibitem [{\citenamefont {Kataoka}\ \emph {et~al.}(2017)\citenamefont
  {Kataoka}, \citenamefont {Fletcher},\ and\ \citenamefont
  {Johnson}}]{kataoka2017timeResolvedWPdetection}%
  \BibitemOpen
  \bibfield  {author} {\bibinfo {author} {\bibfnamefont {M.}~\bibnamefont
  {Kataoka}}, \bibinfo {author} {\bibfnamefont {J.~D.}\ \bibnamefont
  {Fletcher}},\ and\ \bibinfo {author} {\bibfnamefont {N.}~\bibnamefont
  {Johnson}},\ }\href {https://doi.org/10.1002/pssb.201600547} {\bibfield
  {journal} {\bibinfo  {journal} {Phys. Status Solidi B}\ }\textbf {\bibinfo
  {volume} {254}},\ \bibinfo {pages} {1600547} (\bibinfo {year}
  {2017})}\BibitemShut {NoStop}%
\bibitem [{\citenamefont {Johnson}\ \emph {et~al.}(2017)\citenamefont
  {Johnson}, \citenamefont {Fletcher}, \citenamefont {Humphreys}, \citenamefont
  {See}, \citenamefont {Griffiths}, \citenamefont {Jones}, \citenamefont
  {Farrer}, \citenamefont {Ritchie}, \citenamefont {Pepper}, \citenamefont
  {Jannsen},\ and\ \citenamefont
  {Kataoka}}]{johnson2016ultrafastWavePackSampling}%
  \BibitemOpen
  \bibfield  {author} {\bibinfo {author} {\bibfnamefont {N.}~\bibnamefont
  {Johnson}}, \bibinfo {author} {\bibfnamefont {J.~D.}\ \bibnamefont
  {Fletcher}}, \bibinfo {author} {\bibfnamefont {D.~A.}\ \bibnamefont
  {Humphreys}}, \bibinfo {author} {\bibfnamefont {P.}~\bibnamefont {See}},
  \bibinfo {author} {\bibfnamefont {J.~P.}\ \bibnamefont {Griffiths}}, \bibinfo
  {author} {\bibfnamefont {G.~A.~C.}\ \bibnamefont {Jones}}, \bibinfo {author}
  {\bibfnamefont {I.}~\bibnamefont {Farrer}}, \bibinfo {author} {\bibfnamefont
  {D.~A.}\ \bibnamefont {Ritchie}}, \bibinfo {author} {\bibfnamefont
  {M.}~\bibnamefont {Pepper}}, \bibinfo {author} {\bibfnamefont {T.~J. B.~M.}\
  \bibnamefont {Jannsen}},\ and\ \bibinfo {author} {\bibfnamefont
  {M.}~\bibnamefont {Kataoka}},\ }\href {https://doi.org/10.1063/1.4978388}
  {\bibfield  {journal} {\bibinfo  {journal} {Appl. Phys. Lett.}\ }\textbf
  {\bibinfo {volume} {110}},\ \bibinfo {pages} {102105} (\bibinfo {year}
  {2017})}\BibitemShut {NoStop}%
\bibitem [{\citenamefont {Rössler}\ \emph {et~al.}(2010)\citenamefont
  {Rössler}, \citenamefont {Küng}, \citenamefont {Dröscher}, \citenamefont
  {Choi}, \citenamefont {Ihn}, \citenamefont {Ensslin},\ and\ \citenamefont
  {Beck}}]{roessler2010hybridQDchargeDet}%
  \BibitemOpen
  \bibfield  {author} {\bibinfo {author} {\bibfnamefont {C.}~\bibnamefont
  {Rössler}}, \bibinfo {author} {\bibfnamefont {B.}~\bibnamefont {Küng}},
  \bibinfo {author} {\bibfnamefont {S.}~\bibnamefont {Dröscher}}, \bibinfo
  {author} {\bibfnamefont {T.}~\bibnamefont {Choi}}, \bibinfo {author}
  {\bibfnamefont {T.}~\bibnamefont {Ihn}}, \bibinfo {author} {\bibfnamefont
  {K.}~\bibnamefont {Ensslin}},\ and\ \bibinfo {author} {\bibfnamefont
  {M.}~\bibnamefont {Beck}},\ }\href {https://doi.org/10.1063/1.3501977}
  {\bibfield  {journal} {\bibinfo  {journal} {Appl. Phys. Lett.}\ }\textbf
  {\bibinfo {volume} {97}},\ \bibinfo {pages} {152109} (\bibinfo {year}
  {2010})}\BibitemShut {NoStop}%
\bibitem [{\citenamefont {Takada}\ \emph {et~al.}(2019)\citenamefont {Takada},
  \citenamefont {Edlbauer}, \citenamefont {Lepage}, \citenamefont {Wang},
  \citenamefont {Mortemousque}, \citenamefont {Georgiou}, \citenamefont
  {Barnes}, \citenamefont {Ford}, \citenamefont {Yuan}, \citenamefont {Santos},
  \citenamefont {Waintal}, \citenamefont {Ludwig}, \citenamefont {Wieck},
  \citenamefont {Urdampilleta}, \citenamefont {Meunier},\ and\ \citenamefont
  {Bäuerle}}]{takada2019sawBeamsplitter}%
  \BibitemOpen
  \bibfield  {author} {\bibinfo {author} {\bibfnamefont {S.}~\bibnamefont
  {Takada}}, \bibinfo {author} {\bibfnamefont {H.}~\bibnamefont {Edlbauer}},
  \bibinfo {author} {\bibfnamefont {H.~V.}\ \bibnamefont {Lepage}}, \bibinfo
  {author} {\bibfnamefont {J.}~\bibnamefont {Wang}}, \bibinfo {author}
  {\bibfnamefont {P.-A.}\ \bibnamefont {Mortemousque}}, \bibinfo {author}
  {\bibfnamefont {G.}~\bibnamefont {Georgiou}}, \bibinfo {author}
  {\bibfnamefont {C.~H.~W.}\ \bibnamefont {Barnes}}, \bibinfo {author}
  {\bibfnamefont {C.~J.~B.}\ \bibnamefont {Ford}}, \bibinfo {author}
  {\bibfnamefont {M.}~\bibnamefont {Yuan}}, \bibinfo {author} {\bibfnamefont
  {P.~V.}\ \bibnamefont {Santos}}, \bibinfo {author} {\bibfnamefont
  {X.}~\bibnamefont {Waintal}}, \bibinfo {author} {\bibfnamefont
  {A.}~\bibnamefont {Ludwig}}, \bibinfo {author} {\bibfnamefont {A.~D.}\
  \bibnamefont {Wieck}}, \bibinfo {author} {\bibfnamefont {M.}~\bibnamefont
  {Urdampilleta}}, \bibinfo {author} {\bibfnamefont {T.}~\bibnamefont
  {Meunier}},\ and\ \bibinfo {author} {\bibfnamefont {C.}~\bibnamefont
  {Bäuerle}},\ }\href {https://doi.org/10.1038/s41467-019-12514-w} {\bibfield
  {journal} {\bibinfo  {journal} {Nat. Commun.}\ }\textbf {\bibinfo {volume}
  {10}},\ \bibinfo {pages} {4557} (\bibinfo {year} {2019})}\BibitemShut
  {NoStop}%
\bibitem [{fn3()}]{fn34}%
  \BibitemOpen
  \href@noop {} {}\bibinfo {note} {Leftmost "$P_1$"-markers in (a) represent
  insufficient statistics (not non-zero leveling), $P_1$ leveling at $10^{-3}$
  in (b) might result from complex RF excitation occasionally producing
  even-higher-energy electrons.}\BibitemShut {Stop}%
\bibitem [{\citenamefont {Kashcheyevs}\ and\ \citenamefont
  {Samuelsson}(2017)}]{slava2017classQuantCrossover}%
  \BibitemOpen
  \bibfield  {author} {\bibinfo {author} {\bibfnamefont {V.}~\bibnamefont
  {Kashcheyevs}}\ and\ \bibinfo {author} {\bibfnamefont {P.}~\bibnamefont
  {Samuelsson}},\ }\href {https://doi.org/10.1103/PhysRevB.95.245424}
  {\bibfield  {journal} {\bibinfo  {journal} {Phys. Rev. B}\ }\textbf {\bibinfo
  {volume} {95}},\ \bibinfo {pages} {245424} (\bibinfo {year}
  {2017})}\BibitemShut {NoStop}%
\bibitem [{\citenamefont {Fletcher}\ \emph {et~al.}(2019)\citenamefont
  {Fletcher}, \citenamefont {Johnson}, \citenamefont {Locane}, \citenamefont
  {See}, \citenamefont {Griffiths}, \citenamefont {Farrer}, \citenamefont
  {Ritchie}, \citenamefont {Brouwer}, \citenamefont {Kashcheyevs},\ and\
  \citenamefont {Kataoka}}]{fletcher2019tomography}%
  \BibitemOpen
  \bibfield  {author} {\bibinfo {author} {\bibfnamefont {J.~D.}\ \bibnamefont
  {Fletcher}}, \bibinfo {author} {\bibfnamefont {N.}~\bibnamefont {Johnson}},
  \bibinfo {author} {\bibfnamefont {E.}~\bibnamefont {Locane}}, \bibinfo
  {author} {\bibfnamefont {P.}~\bibnamefont {See}}, \bibinfo {author}
  {\bibfnamefont {J.~P.}\ \bibnamefont {Griffiths}}, \bibinfo {author}
  {\bibfnamefont {I.}~\bibnamefont {Farrer}}, \bibinfo {author} {\bibfnamefont
  {D.~A.}\ \bibnamefont {Ritchie}}, \bibinfo {author} {\bibfnamefont {P.~W.}\
  \bibnamefont {Brouwer}}, \bibinfo {author} {\bibfnamefont {V.}~\bibnamefont
  {Kashcheyevs}},\ and\ \bibinfo {author} {\bibfnamefont {M.}~\bibnamefont
  {Kataoka}},\ }\href {https://doi.org/10.1038/s41467-019-13222-1} {\bibfield
  {journal} {\bibinfo  {journal} {Nat. Commun.}\ }\textbf {\bibinfo {volume}
  {10}},\ \bibinfo {pages} {5298} (\bibinfo {year} {2019})}\BibitemShut
  {NoStop}%
\bibitem [{fn2()}]{fn2}%
  \BibitemOpen
  \href@noop {} {}\bibinfo {note} {Histogram evaluation in this region is
  complicated due to electron tunneling \textit{simultaneously} to data
  acquisition. This significantly reduces the fidelity but does not challenge
  the existence of the feature itself.}\BibitemShut {Stop}%
\bibitem [{fn1()}]{fn1}%
  \BibitemOpen
  \href@noop {} {}\bibinfo {note} {These events are not a parasitic effect,
  e.g. by coupling of the SEP excitation onto \nl{}. $P_{-2}=P_{-1}=0$ is
  restored if electron emission is suppressed (by raising \pd{} or by turning
  off one or both SEP excitations).}\BibitemShut {Stop}%
\bibitem [{\citenamefont {Taubert}\ \emph {et~al.}(2010)\citenamefont
  {Taubert}, \citenamefont {Schinner}, \citenamefont {Tranitz}, \citenamefont
  {Wegscheider}, \citenamefont {Tomaras}, \citenamefont {Kehrein},\ and\
  \citenamefont {Ludwig}}]{taubert2010avalancheAmplifier}%
  \BibitemOpen
  \bibfield  {author} {\bibinfo {author} {\bibfnamefont {D.}~\bibnamefont
  {Taubert}}, \bibinfo {author} {\bibfnamefont {G.~J.}\ \bibnamefont
  {Schinner}}, \bibinfo {author} {\bibfnamefont {H.~P.}\ \bibnamefont
  {Tranitz}}, \bibinfo {author} {\bibfnamefont {W.}~\bibnamefont
  {Wegscheider}}, \bibinfo {author} {\bibfnamefont {C.}~\bibnamefont
  {Tomaras}}, \bibinfo {author} {\bibfnamefont {S.}~\bibnamefont {Kehrein}},\
  and\ \bibinfo {author} {\bibfnamefont {S.}~\bibnamefont {Ludwig}},\ }\href
  {https://doi.org/10.1103/PhysRevB.82.161416} {\bibfield  {journal} {\bibinfo
  {journal} {Phys. Rev. B}\ }\textbf {\bibinfo {volume} {82}},\ \bibinfo
  {pages} {161416(R)} (\bibinfo {year} {2010})}\BibitemShut {NoStop}%
\bibitem [{\citenamefont {Harbusch}\ \emph {et~al.}(2010)\citenamefont
  {Harbusch}, \citenamefont {Taubert}, \citenamefont {Tranitz}, \citenamefont
  {Wegscheider},\ and\ \citenamefont
  {Ludwig}}]{harbusch2010detectorBackaction}%
  \BibitemOpen
  \bibfield  {author} {\bibinfo {author} {\bibfnamefont {D.}~\bibnamefont
  {Harbusch}}, \bibinfo {author} {\bibfnamefont {D.}~\bibnamefont {Taubert}},
  \bibinfo {author} {\bibfnamefont {H.~P.}\ \bibnamefont {Tranitz}}, \bibinfo
  {author} {\bibfnamefont {W.}~\bibnamefont {Wegscheider}},\ and\ \bibinfo
  {author} {\bibfnamefont {S.}~\bibnamefont {Ludwig}},\ }\href
  {https://doi.org/10.1103/PhysRevLett.104.196801} {\bibfield  {journal}
  {\bibinfo  {journal} {Phys. Rev. Lett.}\ }\textbf {\bibinfo {volume} {104}},\
  \bibinfo {pages} {196801} (\bibinfo {year} {2010})}\BibitemShut {NoStop}%
\bibitem [{\citenamefont {Altimiras}\ \emph {et~al.}(2010)\citenamefont
  {Altimiras}, \citenamefont {le~Sueur}, \citenamefont {Gennser}, \citenamefont
  {Cavanna}, \citenamefont {Mailly},\ and\ \citenamefont
  {Pierre}}]{altimiras2010tuningEnergyRelax}%
  \BibitemOpen
  \bibfield  {author} {\bibinfo {author} {\bibfnamefont {C.}~\bibnamefont
  {Altimiras}}, \bibinfo {author} {\bibfnamefont {H.}~\bibnamefont {le~Sueur}},
  \bibinfo {author} {\bibfnamefont {U.}~\bibnamefont {Gennser}}, \bibinfo
  {author} {\bibfnamefont {A.}~\bibnamefont {Cavanna}}, \bibinfo {author}
  {\bibfnamefont {D.}~\bibnamefont {Mailly}},\ and\ \bibinfo {author}
  {\bibfnamefont {F.}~\bibnamefont {Pierre}},\ }\href
  {https://doi.org/10.1103/PhysRevLett.105.226804} {\bibfield  {journal}
  {\bibinfo  {journal} {Phys. Rev. Lett.}\ }\textbf {\bibinfo {volume} {105}},\
  \bibinfo {pages} {226804} (\bibinfo {year} {2010})}\BibitemShut {NoStop}%
\bibitem [{\citenamefont {Locane}\ \emph {et~al.}(2019)\citenamefont {Locane},
  \citenamefont {Brouwer},\ and\ \citenamefont
  {Kashcheyevs}}]{locane2019dynScattSEWP}%
  \BibitemOpen
  \bibfield  {author} {\bibinfo {author} {\bibfnamefont {E.}~\bibnamefont
  {Locane}}, \bibinfo {author} {\bibfnamefont {P.~W.}\ \bibnamefont
  {Brouwer}},\ and\ \bibinfo {author} {\bibfnamefont {V.}~\bibnamefont
  {Kashcheyevs}},\ }\href {https://doi.org/10.1088/1367-2630/ab3fbb} {\bibfield
   {journal} {\bibinfo  {journal} {New J. Phys.}\ }\textbf {\bibinfo {volume}
  {21}},\ \bibinfo {pages} {093042} (\bibinfo {year} {2019})}\BibitemShut
  {NoStop}%
\end{thebibliography}%

\end{document}